\documentclass[10pt,final,journal,twocolumn,twoside,a4paper]{IEEEtran}

\usepackage{epstopdf}
\usepackage{times,amsmath,epsfig}
\usepackage{amssymb}
\usepackage{subfigure}
\usepackage{cite}

\usepackage[table]{xcolor}
\usepackage{todonotes}

\usepackage{algorithmic}
\usepackage{algorithm}
\usepackage[font=small,format=plain,up,up]{caption}
\newcommand{\EE}{{\mathbb E}}
\newcommand{\ctk}{\Delta_k}
\newcommand{\tildectk}{\tilde{\Delta}_k}
\newcommand{\clip}{\boldsymbol{\phi}}
\newtheorem{mytheorem}{\hspace{-11pt}\bf Theorem}

\psfull

\begin{document}

\title{An MDP Model for Censoring in Harvesting Sensors: Optimal and Approximated Solutions}

\author{\vspace*{-0.1cm}
 Jesus~Fernandez-Bes,~\IEEEmembership{Student Member,~IEEE}, Jes\'us~Cid-Sueiro, Antonio~ G.~ Marques,~\IEEEmembership{Senior Member,~IEEE}
\thanks{J.~Fernandez-Bes and J.~Cid-Sueiro are with the Dept. of Teor\'ia de la Se\~{n}al y Comunicaciones, {Univ.} Carlos III de Madrid, Avda. de la Universidad 30, Legan\'es, {28911}, Madrid, Spain. E-mails: (\{jesusfbes,jcid\}@tsc.uc3m.es).}
\thanks{Antonio G. Marques is with the Dept. of Teor\'ia de la Se\~{n}al y Comunicaciones, Universidad Rey Juan Carlos de Madrid, Camino del Molino s/n, Fuenlabrada 28943, Madrid, Spain. E-mail: (antonio.garcia.marques@urjc.es)}
\thanks{This work was supported by the Spanish MECS FPU fellowship programme, Spanish Ministry of Economics grants Nos. TEC2011-22480, TEC2013-41604-R, PRI-PIBIN-2011-1266, and EU FP7-ICT-2011-9 TUCAN3G.}}

\markboth{IEEE JOURNAL ON SELECTED AREAS IN COMMUNICATIONS (submitted, date )}
{FERNANDEZ-BES \MakeLowercase{\textit{et al.}}: An MDP Model for Censoring in Harvesting Sensors: Optimal and Approximated Solutions}
\maketitle


\begin{abstract}%

In this paper we propose a novel censoring policy for energy-efficient transmissions in energy-harvesting sensors. {The problem is formulated as an infinite-horizon Markov Decision Process (MDP). The objective to be optimized is the expected sum of the importance (utility) of all transmitted messages.} Assuming that {such importance} can be evaluated at the transmitting node, we show that, under certain conditions on the battery model, the optimal censoring policy is a threshold function {on the importance value}. Specifically, messages are transmitted only if their importance is above a threshold whose value depends on the battery level. {Exploiting this property, we propose a model-based stochastic scheme that approximates the optimal solution, with less computational complexity and faster convergence speed than a conventional Q-learning algorithm}. Numerical experiments in single-hop and multi-hop networks confirm the analytical advantages of the proposed scheme.
\end{abstract}


\section{Introduction}
\label{sec:introduction}

Efficient management of energy resources is essential to operate Wireless Sensors Networks (WSN) equipped with finite-size batteries and energy-harvesting devices \cite{gunduz2014designing}. Numerous works have designed energy-saving strategies that account for the limited and stochastic nature of the harvested energy. Many of them were aimed at solving general communication problems such as utility-based cross-layer design \cite{gatzianas2010control}, power allocation \cite{tutuncuoglu2012communicating,ho2012optimal}, or rate adaptation \cite{khairnar2011power}. At the same time, in the field of WSN there has been a growing interest in strategies that take into account the importance of the information to be transmitted for the application at hand. {The ``importance value'' can be, for instance, the traffic priority of the routing protocol, the deviation from the mean in distributed estimation \cite{imer2010optimal}, or the likelihood ratio in decentralized detection \cite{appadwedula2008decentralized}.} Such strategies, sometimes referred to as \emph{selective communication} \cite{ArroyoVallesEtAl09} or sensor \emph{censoring} strategies \cite{appadwedula2008decentralized}, assume that nodes can evaluate/quantify the importance of the current message and use it to decide whether transmitting or censoring it. To make the decision, additional parameters such as the cost of the communication, the confidence that the message will arrive to its destination, or the \emph{available energy resources}, should also be taken into account. That is precisely the objective of this work: designing optimal censoring policies for an energy-harvesting scenario.

Since the aforementioned parameters are correlated across time, decisions, which are made sequentially, should be designed to optimize the long-term behavior of the system (for example, by maximizing the aggregated importance of all messages transmitted by the WSN). From an algorithmic viewpoint, the design of such censoring policies is a Dynamic Programming (DP) problem that, under certain assumptions, can be modeled as a Markov Decision Process (MDP). The current ``transmit or censor'' decision changes the amount of energy stored in the battery and, therefore, has an impact not only on the current battery state, but also on future ones. Therefore, efficient policies have to balance the benefits of an immediate reward with the (expected) impact of each decision on future costs/rewards. In general, the resulting DP problems are difficult to solve and {approximate solutions} are often required \cite{bertsekasDP,powell_book}.

{Assuming a ``predictable'' setting where one knows not only the past, but also the future values of the state information (e.g., energy to be harvested in the future instants), optimal off-line decision policies can be applied \cite{tutuncuoglu2012optimum,yang2012optimal}. Although, such schemes serve as a benchmark, they do not cope well with many practical scenarios, where energy and packet-arrival processes are not known in advance. Consequently, {many works in the} literature are MDP based.} In \cite{Lei09} a policy iteration algorithm \cite{bertsekasDP} is used to estimate the decision policy maximizing the long-term average reward, for a dual recharge/replace battery harvesting model with unitary transmission costs. The scheme in \cite{michelusi2012optimal} is also based on unitary costs and an average reward optimization. This work and its later extension in \cite{michelusi2013transmission} show that (a) the optimal transmission policy applies a threshold over the importance values that is a decreasing function of the available energy, and (b) a \emph{balanced policy} is close to optimal. {The balanced policy is a simple scheme, also analyzed in this paper, that takes into account the long-term distribution of the energy harvested and ignores the \emph{instantaneous} battery level \cite{michelusi2013transmission,ArroyoVallesEtAl11}.}

The main drawback of these approaches is that they need to know the distributions of the energy and packet arrival processes, which may not be available in practical scenarios. In such cases, nodes have to be able to learn whatever information is needed in real time. One approach to handle this problem, recently proposed in \cite{Blasco2013}, is to apply Q-learning (a stochastic method widely used in the context of reinforcement learning \cite{bertsekasDP,powell_book}) to solve the DP problem. But this approach has its own problems. {Since} Q-learning is a model-free algorithm that tries to estimate the value function in a non-parametric manner, it can result in a slow convergence and high computational complexity when the size of the state space grows {(the problem is even more severe if the state space is continuous)}.

{In this paper, we propose a \emph{model-based stochastic approximation algorithm} to solve the previous difficulties. More specifically, we show that, under reasonable assumptions on the battery dynamics, the optimal censoring policy is a threshold function on the importance value. An ad-hoc stochastic approximation algorithm exploiting this property is proposed that, compared with a conventional Q-learning algorithm, is more efficient in terms of computational complexity, memory requirements and convergence speed.}

A similar approach for the case of censoring sensors with non-rechargeable batteries was proposed for single-hop communications \cite{ArroyoVallesEtAl09}, and for the optimization of multi-hop networks\cite{ArroyoVallesEtAl11,fernandez2011cooperative}. Compared to our previous work in scenarios with \emph{non-rechargeable} batteries \cite{ArroyoVallesEtAl09,ArroyoVallesEtAl11}, the results in this paper reveal that the harvesting scenario is substantially different. While, in a non-rechargeable case, a censoring policy discarding messages whose importance is below a \emph{fixed threshold} is quasi-optimal, censoring policies based on \emph{energy-dependent thresholds} are significantly more efficient in harvesting sensors.

Our experiments, based on a battery model that copes with more general stochastic transmission and reception costs, and also for stochastic harvesting patterns, show that energy-dependent transmission policies can be more efficient than balanced policies. Our use of a penalized discount with infinite horizon \cite{bertsekasDP} is also more appropriate to cope with non-stationary environments. As in almost every approach in the literature, our MDP model optimizes the performance of each node separately, which is guaranteed to be globally optimal only for single-hop scenarios. However, by using a success index \cite{ArroyoVallesEtAl11} to model the behavior of neighboring nodes, MDP models at different nodes get coupled and, as a result, a network of MDP-based sensors can significantly outperform networks where nodes implement balanced policies.

The remaining of the paper is organized as follows. Section \ref{S:Sensor_model} describes {the WSN model}. Section \ref{S:Opt_sel_fw} states the main theoretical results. Section \ref{Sec.SingleNode} considers different simplifying assumptions to gain insights on the optimal solution. Section \ref{S:Appr_sche} develops low-complexity \emph{stochastic} schemes that approximate the optimal solution and are robust to {non-stationarities}. Section \ref{S:Simulations} evaluates the proposed schemes using numerical simulations. Conclusions in Section \ref{S:Conclusions} close the paper.

\section{System model}\label{S:Sensor_model}

{In this section, we introduce notation, explain the mathematical model that describes the dynamics of our system under a censoring policy,} and formulate the objective to be optimized.

The model is defined by four main components: a) a set of state variables, b) a set of possible actions, c) a probabilistic model of the state dynamics (that describes how future states depend on the current state and the actions taken), and d) a reward model (that describes the immediate reward obtained when some action is taken at a given state). As explained in the introduction, since we are interested in maximizing the long-term reward and current actions have an impact into the future states, our problem will fall into the DP framework. Moreover, because the state dynamics are assumed Markovian, the problem will be modeled as an MDP.

\subsection{State vector}


{Consider a node that receives a sequence of requests to transmit different messages.} The messages can be received from another node or generated from local  measurements. The state of the node will be characterized by two variables
\begin{itemize}
 \item $e_k$: the battery level at step (slot) $k$. It reflects the ``internal state'' of the node,
 \item $x_k$: the importance of the message to be sent at step $k$.
\end{itemize}

Following the typical terminology in MDP models, the state vector of the node is defined as ${\bf s}_k = (e_k,x_k)$; i.e., the state vector contains all and only the information that is \emph{available at the node} to make a decision at time $k$. The set of all possible states is denoted as ${\mathcal S}$.

{To facilitate exposition, $k$ is considered an \emph{epoch} or slot index, which starts when the node has to decide whether to censor or transmit a message (either received from one of its neighbors or generated from its sensing devices) and ends when the next message is received. Besides transmitting or censoring the message, during each epoch the node (eventually) collects some energy from the environment. This approach, which is used by many authors, implies that the actual duration of each slot $k$ is stochastic. Clearly, the results in the paper also hold true if the system operates with a constant sampling period (the only modification required is to set $x_k=0$ for the time instants $k$ where no message has been received).}

{Note that, besides $e_k$ and $x_k$, the node could use additional information to make decisions. This information can be also local (the packet length, the state of the communication channel) or pertain to other (neighboring) nodes. Additional local information can be easily incorporated into the formulation, provided that the state dynamics are similar to those of $e_k$ and $x_k$; see, e.g., \cite{ArroyoVallesEtAl11}. Incorporating information about the state or the eventual actions of neighboring nodes (e.g., battery levels, or information about their censoring policy) will lead to a better network operating point, although it raises issues such as the accuracy and the cost of acquiring the non-local information (exchange of information requires, for example, additional energy consumption). Since the paper investigates the design of separate (per-node) censoring policies, we will focus on local information. As we will see in Section \ref{S:Opt_sel_fw}, the success index variable defined in Section \ref{S:Rewards} could be used as a means to couple the decisions across the network; see \cite{ArroyoVallesEtAl11} for details.}

\subsection{Actions and policies}

At each time epoch $k$, the sensor node must take an action (decision) $a_k$ about sending the current message ($a_k=1$), or censoring it ($a_k=0$). A forwarding policy $\pi = \{a_1,a_2,\ldots \}$ at a given node is a sequence of decision rules, which are functions of the state vector; i.e.,
\begin{equation}
a_k = \pi_k({\bf s}_k) = \pi_k(e_k,x_k).
\label{Eg.General0Trasmitter}
\end{equation}

\subsection{State dynamics}

Next, we describe the model for the stochastic processes $e_k$ and {$x_k$} {that form the state vector.}

The energy consumed at each time epoch depends on the taken action. { Let cost $\hat{c}_{k}$ denote the energy consumed by the node and $b_k$ the amount of energy (if any) harvested by the node since the last action $a_{k-1}$. Then, $e_{k+1}$ can be written recursively as $e_{k+1} = \clip_B\big(e_k - \hat{c}_k + b_k\big)$}, where $\clip_B(e)= \max(0,\min(e,B))$ is a clipping (projection) function that guarantees that the energy stored in the battery is never negative, nor exceeds its maximum capacity $B$.\footnote{Similar models are used in related works \cite{gunduz2014designing}. For example, \cite{michelusi2012optimal} use a slightly different model, {$e_{k+1} = \min\{\max\{e_k - \hat{c}_k, 0 \}  + b_k, B\}$}, which assumes that the energy recharge happens at the end of the decision slot. Using this alternative model does not state special difficulties, and would not change the qualitative analysis in this paper.}

Cost {$\hat{c}_k$} may include the cost of data sensing (if the sensor is the source of the message), the cost of data \emph{reception} (when data come from other nodes), the cost of idle periods, or whatever other costs incurred since the last action. {When $a_k=1$, $\hat{c}_{k}$  includes the previous costs plus the cost of \emph{transmitting} the message.} This statistical model allows us to deal with a broad range of scenarios: stochastic packet arrivals, communications over fading channels, packet losses, automatic repeat request (ARQ) schemes, {to name a few}. In those networks, the energy consumption during node communications can vary depending on the amount of retransmissions required for a successful packet arrival. The range of values and statistical model for $b_k$ depend on both the type of harvesting device and the source of energy considered \cite{Kansal2007}. {To simplify notation, we define $c_k=\hat{c}_{k}-b_k$, so that battery dynamics can be rewritten in a more compact form as
\begin{equation}
e_{k+1} = \clip_B\left(e_k - c_k \right).
\label{Eq:EnergyDynamics}
\end{equation}}
{Note that high values of harvested energy can render $c_k$ negative.}





\subsection{Rewards}
\label{S:Rewards}

The {\em reward} at time $k$ is given by
\begin{equation}
r_k = a_k w_k x_k,
\label{EqReward}
\end{equation}
where $w_k \in \{0,1\}$ denotes the {\em success index} (a binary variable taking value 1 if the transmission is successful, and zero otherwise). {Thus, the reward $r_k$ that each node receives is a positive value $x_k$ if and only if it decides to transmit the message ($a_k=1$) and the transmission is successful ($w_k=1$). Otherwise, the reward is zero.}

The meaning of the success index {$w_k$} depends on the application scenario. In general, a reasonable choice is to set $w_k=1$ if and only if the message is properly received at its final destination. However, in multi-hop networks, the information about the reception of messages at the sink node may not be available to all nodes along the route. In such a case, other (suboptimal) choices for $w_k$ are possible. For instance, in \cite{ArroyoVallesEtAl11} it is shown that setting $w_k=1$ if the neighboring node forwards the transmitted message can be nearly as efficient as using the actual information from the sink. Another (simpler) way to decouple the decisions between nodes is just setting $w_k=1$ when the node is able to transmit a message, as proposed in \cite{ArroyoVallesEtAl09}. This choice is optimal in single-hop networks with star topology. {In any case, the optimal policy in Section \ref{S:Opt_sel_fw} will demonstrate that the optimal action depends on $w_k$ or, to be more precise, on the knowledge of $w_k$ available at the agent making the decision.}


\subsection{Problem formulation}

Our transmission policies will be designed so that the expected aggregate reward is maximized. {Following a standard approach in DP, the ``so-called'' discount factor $0<\gamma<1$ is considered \cite{bertsekasDP} and, based on it, the expected aggregate reward is defined as}
\begin{align}
V_{\pi}({\bf s}) = \EE\left\{\sum_{k=0}^{\infty} \gamma^k r_k | {\bf s}_0={\bf s}\right\} 
                 = \EE\left\{\sum_{k=0}^{\infty} \gamma^k a_k w_k x_k|{\bf s}_0={\bf s}\right\}.
\label{EqSOrder2}
\end{align}
The optimal transmission policy is then
\begin{equation}
\pi^* = \arg\max_{\pi} V_{\pi}.
\end{equation}
Note that only messages {\em successfully} transmitted by the nodes are relevant in \eqref{EqSOrder2}.
Eq. \eqref{EqSOrder2} states a DP problem with infinite horizon (because all future time instants are considered) and discounted cost (due to the presence of $\gamma$ which penalizes future rewards exponentially) \cite{bertsekasDP}. Indeed, $V_{\pi}({\bf s}) $ is typically referred to as either value function or reward-to-go function. Mathematically, the presence of $\gamma$ eases the existence of a stationary policy that optimizes \eqref{EqSOrder2}; see, e.g., \cite{bertsekasDP}. Additional details will be given in the ensuing section.


\section{Optimal stationary policy}\label{S:Opt_sel_fw}

\begin{figure}[t]
\centering
\includegraphics[clip,width=1\columnwidth]{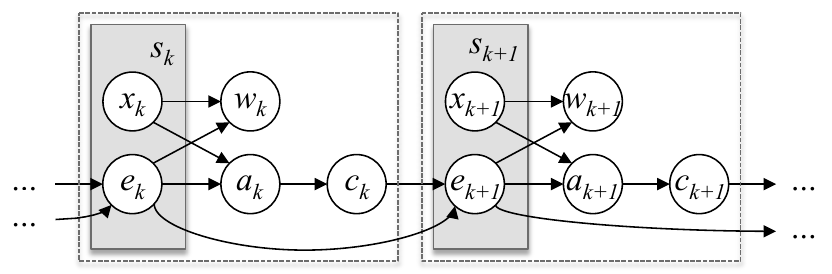}
\caption[Graphical Model]{{A graphical model relating the main variables in the MDP.}}
\label{fig:MDPGM}
\end{figure}

This section is devoted to design stationary solutions that solve the DP formulated in Section II.E. Since the objective in \eqref{EqSOrder2} depends on the stochastic processes $x_k$, {$c_k$}, and $w_k$, assumptions on the stationarity of such processes are required.
{The relationships among the main variables in the MDP are represented in the graphical model in Fig. \ref{fig:MDPGM}. Arrows in this model encode direct causal relationships between variables: the action is a function of the state, the success index depends on the state only (and also on $a_k$, which is a deterministic function of the state), the energy consumed or harvested depends on the action taken, and the energy at the next state depend on the current battery level and the energy consumed or harvested at time $k$. The model assumptions underlying the graphical model representation and that will be used in our analysis, are the following}:\\
\noindent as1) the process $x_k \ge 0$ is independent, identically distributed (i.i.d.) {and independent of $e_k$};\\
\noindent as2) $c_k$ is independent of $x_k$, $e_k$ and all its previous history, given the action, $a_k$, and $p(c_k|a_k)$ does not depend on $k$;\\
\noindent as3) $w_k$ is independent of all its previous history, given $e_k$ and $x_k$, and $p(w_k|e_k,x_k)$ does not depend on $k$.\\
\noindent{Some independence assumptions may be oversimplifying for some applications: in particular, the independence of the importance values can be non realistic in scenarios where consecutive sensor measurements are correlated. Also, the harvested energy can be time-correlated when it depends on environmental variables that span over several epochs} (on the other hand, the energy harvested by wind sensors is oftentimes modeled as i.i.d. \cite{CartaRV09}). Nonetheless, it is worth mentioning that: {i) incorporating time-dependence into our model (while preserving Markovianity) does not state special difficulties, though it would imply some extra computational load and memory requirements, ii) the independence of $c_k$ with respect to} $x_k$ or $e_k$ can be relaxed without entailing a big penalty in terms of complexity \cite{ArroyoVallesEtAl11}; iii) due to the presence of the discount factor $\gamma$, stationarity can be relaxed to short-term stationarity (more specific comments will be provided in this section after presenting the optimal solution); and iv) the stochastic schemes proposed in Section V will be able to handle non-stationarities.

Under the previous assumptions and due to the recursive definition of $e_k$ given by \eqref{Eq:EnergyDynamics}, the state dynamics are Markovian. Hence, the tuple $({\mathcal S},{\mathcal A},P,r)$, where ${\mathcal S}$ is the set of states, ${\mathcal A}=\{0,1\}$ is the \emph{finite} set of possible decisions (actions), $P$ is the transition probability measure that can be expressed as $p({\bf s}_{k+1}| {\bf s}_k, a_k) = p(e_{k+1}|e_k,a_k)p(x_{k+1})$, and $r$ is the instantaneous reward function, constitutes an MDP. As a result the existence of a Markovian and stationary optimal policy $\pi^*$ is guaranteed \cite{Puterman:05,bertsekasDP}.

Clearly, the value function associated with this policy needs to satisfy the Bellman's optimality equation \cite{bertsekasDP}
\begin{align}
V_{\pi^\ast}({\bf s}) = \max_{a\in\{0,1\}}
      \EE\{r_k + \gamma V_{\pi^*}({\bf s}_{k+1})|a_k=a,{\bf s}_k={\bf s} \},
\label{EqBellman}
\end{align}
which can be used to obtain the optimal decision rule. This is accomplished by Theorem \ref{Th.General}. All expectations in the following are computed over {$x_k$, $c_k$ and $w_k$} (which are the primary random variables in the model), unless otherwise stated through the conditional operators.

\begin{mytheorem} \label{Th.General}

{\em  Under {as1-as3}, it holds that \eqref{EqSOrder2} is maximized by a stationary policy $a_k = \pi^*({\bf s})$ satisfying
\begin{equation}
a_k = u(W(e_k,x_k) x_k-\mu(e_k)),
\label{Eq.opt.a_k.Stat}
\end{equation}
where $u$ is the Heaviside step function, $W(e,x) = \EE\{w_k | e_k=e,x_k=x\}$ is the success probability and the threshold function $\mu$ is defined recursively through the pair of equations
\begin{align}
\label{Eq.Mu_Stat}
\mu(e)
   = & \gamma \left(   \EE\{\lambda(\clip_B(e-{c_k})) { | a_k=0}\} \right.    \nonumber\\
      &               \left. - \EE\{\lambda(\clip_B(e-{c_k})) {| a_k=1}\} \right),   \\
\lambda(e)
   = & \gamma \EE\{\lambda(\clip_B(e - {c_k})) {| a_k=0} \}     \nonumber\\
      &+ \EE\{(W(e,x_k) x_k- \mu(e))^+ \},
\label{Eq.Lambda_Stat}
\end{align}}
with $(z)^+ = \max\{z,0\}$, for any $z$.

The auxiliary function $\lambda(e)$ represents the expected value function for an initial battery $e_0=e$, i.e.,
\begin{eqnarray}
\lambda(e) = \EE\{V_{\pi^*}({\bf s})|e_0=e\}.
\label{EqDefLambda}
\end{eqnarray}
\end{mytheorem}

\begin{IEEEproof}
See Appendix A.
\end{IEEEproof}

This theorem comes from the direct application of Bellman's equation to a case where some part of the state, $x$, is uncontrollable, and the ``reduced'' value function $\lambda$ only depends on $e$. A discussion on this kind of problems can be found in, e.g., \cite[Chap.~6.1]{bertsekas2011dynamic}. Focusing for now on \eqref{Eq.opt.a_k.Stat}, the theorem establishes that transmission decisions are made by comparing the expected instantaneous reward $W(e_k,x_k) x_k$ with an energy-dependent threshold $\mu(e_k)$. {The threshold quantifies the loss of the future reward associated with the transmission. In other words, the value of $\mu(e_k)$ is the difference between the future reward if $a_k=0$ (and thus transmission energy is preserved) and that if $a_k=1$. Clearly, the expected future reward will be higher if the energy stored in the battery is higher (because more transmissions can be afforded), so that $\lambda(\cdot)$ is an increasing function of $e$ and $\mu(\cdot)$ is always positive. This implies that the optimal policy compares the instantaneous and future rewards and acts accordingly. Moreover, the instantaneous reward depends on the (expected) value of the success index, confirming that the optimal action depends not only on $e_k$ and $x_k$, but also on $w_k$.}

Due to as1-as3, the policy in Theorem 1 is stationary. As a result, the optimal transmission policy (mapping from state variables to actions) does not depend on the specific time instant, but only on the value of the state variables. As mentioned earlier, the presence of $\gamma$ opens the door to deal with non-stationarities as long as the state information is stationary in the short-term. Intuitively, the reason is that $\sum_{k=0}^\infty\gamma^k$ can be safely approximated by $\sum_{k=0}^N\gamma^k$ with $N<\infty$ (for instance, for $\gamma=0.95$ and $N=100$ the error in the approximation is less than $1\%$). Provided that the processes are stationary during at least an interval of length $N$, using the results in Theorem 1 will lead to a small error. In such a case, the transmission policy would need to be recomputed every time the \emph{distribution} of the random processes changes.

Although Theorem \ref{Th.General} holds for any cost and importance distributions, it does not provide a clear intuition on how such distributions influence the optimal policies. Additionally, the resulting equations are difficult to solve (even if the expectations involved can be computed). The remaining of the paper is devoted to handle some of these issues. 
In Section \ref{Sec.SingleNode} we consider several simplifying assumptions that render the theoretical analysis more tractable, so that we can get further insights on the behavior of the optimal solution. Finally, in Section \ref{S:Appr_sche} we design low-complexity stochastic approximations to the analytical schemes that can be applied in general scenarios (not only to those in Section \ref{Sec.SingleNode}).

\section{Analysis of the optimal policy}
\label{Sec.SingleNode}

In this section, we will analyze different aspects of the optimal policies. In the first subsection, we will derive recursive expressions to compute those optimal policies. In the second subsection, we will obtain the steady-state energy distributions and assess their impact on the optimal policies. To facilitate the computation of the optimal schemes, we will consider three simplifying \emph{assumptions}: AS1) Process $w_k$ does not depend on $x_k$. AS2) The energy variables are discretized, so that $e_k$, {$c_k$}  and $B$ are integer values. {As a result, the energy space is approximated by a finite space, but the approximation error can be minimized by choosing the energy resolution $\varepsilon$ small enough, though at the expense of increasing the memory requirements and the computational complexity. Discretization is a widely used approach to deal with continuous-state DPs.} AS3) The success probability can be written as
\begin{align}
W(e) = {P\{c_k \le e | a_k=1 \}}  
   \label{EqErk1}
\end{align}
In words, the transmission is successful if the node has energy enough to transmit the message. This is the case if, for example, the communications are error free. In the presence of path losses, \eqref{EqErk1} also holds if the message is retransmitted until a confirmation is received --the path-loss probability would modify the distribution of the energy cost {$c_{k}$, but not the formal expression in} \eqref{EqErk1}. Alternatively, if retransmissions are finite (or zero) path losses can be accommodated by just multiplying {the right-hand side of  \eqref{EqErk1} by} the packet loss probability. This equation is specially suited to single-hop communications. In multi-hop networks, it is suboptimal because it does not consider whether the message is eventually forwarded through the network up to the sink. Nonetheless, the equation decouples variables across nodes, simplifying the derivation of analytical expressions that can be useful even for scenarios where it entails a loss of optimality.

{Note that these assumptions are only needed to have a simple scenario where the optimal policy can be computed, and consequently get some insights on its structure and performance. Once that goal is achieved, they will be relaxed in the following sections.}

\subsection{Optimal policies for particular cases}


Under assumptions AS1 and AS2, the selective transmitter given by \eqref{Eq.opt.a_k.Stat}, \eqref{Eq.Mu_Stat} and \eqref{Eq.Lambda_Stat} can be described by the following set of discrete equations
\begin{align}
a_k &= u\left(x_k-\frac{\mu(e_k)}{W(e_k)}\right)
\label{Eq.a_kOptStat2}\\
\mu(e) =& \gamma \EE\{\lambda(\phi_B(e-{c)) | a=0}\}   \nonumber\\
             &- \gamma \EE\{\lambda(\phi_B(e-{c)) | a=1}\}
\label{Eq.MuS2}\\
\lambda(e) &= \gamma \EE\{\lambda(\phi_B(e-{c)) | a=0}\}
           + W(e) h\left(\frac{\mu(e)}{W(e)}\right)
\label{Eq.LambdaS2}
\end{align}
where
\begin{equation}
h(\alpha) = \EE\{(x-\alpha)^+\}.
\label{Eq.DefH2}
\end{equation}
Note that since $c_k$ and $a_k$ are stationary, subindex $k$ has been dropped to simplify notation.

Even in this simplified scenario, \eqref{Eq.MuS2} and \eqref{Eq.LambdaS2} cannot be solved analytically, so that neither the ``reduced'' value function nor the transmission threshold can be found in closed form. However, the considered assumptions reduce the size of the state space and thus, facilitate the implementation of iterative methods, such as value iteration and policy iteration \cite{bertsekasDP,powell_book,Puterman:05}. {For example, using Value Iteration and with $l$ denoting an iteration index, the optimal schemes can be found using the iterations}
\begin{align}
\mu_l(e) =& \gamma \EE\{\lambda_{l-1}(\phi_B(e-{c)) | a=0}\}   \nonumber\\
                &- \gamma \EE\{\lambda_{l-1}(\phi_B(e-{c)) | a=1}\}
\label{Eq.Mu_1c}\\
\lambda_l(e) &= \gamma \EE\{\lambda_{l-1}(\phi_B(e-{c)) | a=0}\}
             + W(e) h\left(\frac{\mu_{l}(e)}{W(e)}\right)
\label{Eq.Lambda1c}
\end{align}
{where $\mu_0(e)$ and $\lambda_0(e)$ are the arbitrary initial values.}




To gain some insights, next we solve numerically and analyze the optimal solution for different scenarios with stochastic energy costs. We consider the case when {$\EE\{c | a=0\}<0$}, which implies that nodes can discard messages to recharge batteries during operation and, thus, the lifetime can be extended indefinitely. {To be more meaningful, we will assume a scenario where each decision epoch can be split into a variable number of fixed-duration time slots, and variable $c$ can be decomposed as}
\begin{equation}
{c = n_S \cdot c_I + c_R - b + a \cdot \Delta,}\label{eq:cost_model0}
\end{equation}
{where $n_S$ is the number of time slots since the last decision, $c_I$ is the (stand-by) energy consumed during each time slot}, $c_R$ is the cost of receiving or sensing the current message, $b$ is the amount of battery recharged since the last node decision, {$a$ is the action, and $\Delta$ is the incremental cost of deciding $a=1$. We assume a lossy channel where retransmission trials are repeated until the message is successfully received at destination. Thus,}
\begin{equation}
{\Delta = n_T c_T}\label{eq:cost_model1}
\end{equation}
{where $c_T$ is the cost of each transmission trial and $n_T$ is the number of transmission trials.}

We have simulated a scenario {where $c_R=2$, $c_I=1$ and $n_S$ follow a geometric distribution with mean 2}. We assumed a very poor channel, so that {transmission trials fail with probability $0.4$}. The cost of each transmission trial is set to $4$. {This configuration tries to simulate WSN configurations where the energy cost of transmitting a message is substantially higher than that of sensing or receiving a message.} The amount of battery recharged, $b$, is also stochastic.{ We assume that the amount of battery recharge can be decomposed as $b=\sum_{i=1}^m b_i$ where $b_i$ are i.i.d. variables accounting for the battery recharged at each time slot $i$. During each time slot, the probability of a nonzero battery recharge is $p_b=1/3$, and, when $b_i>0$, $b_i$ is geometrically distributed with mean $m_b$}. Three different values of $m_b$, namely, {5, 10 and 15}, have been explored. For these values, the corresponding values of
{\begin{equation}
\overline{c}_0 = \EE\{c |a=0\}
\end{equation}}
are {-0.1, -3.4 and -6.7}, respectively. Finally, an i.i.d. exponential importance distribution with unit mean was assumed, and $\gamma=0.999$.

\begin{figure}[t]
\centering
\subfigure[]{\includegraphics[trim=6mm 68mm 110mm 60mm, clip,width=0.49\columnwidth]{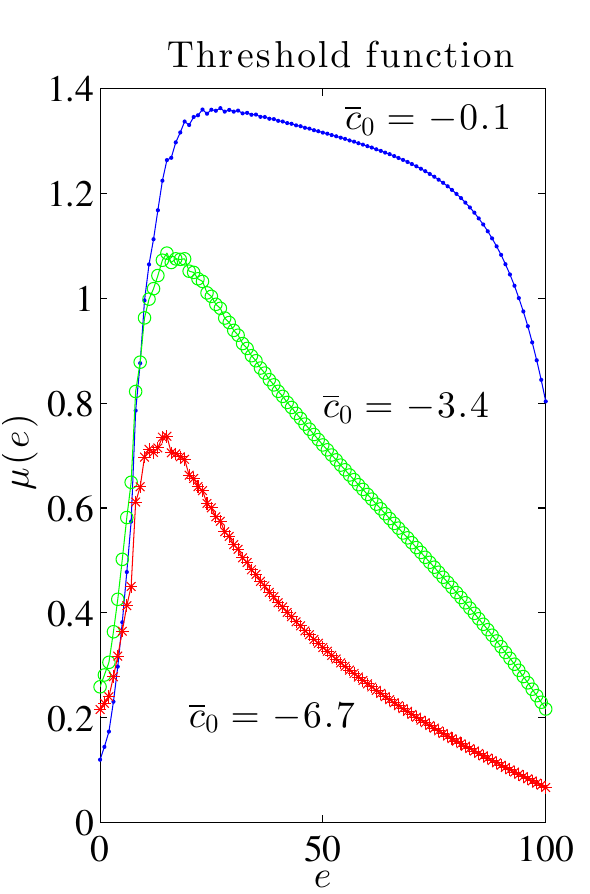}}
\subfigure[]{\includegraphics[trim=6mm 68mm 110mm 60mm, clip,width=0.49\columnwidth]{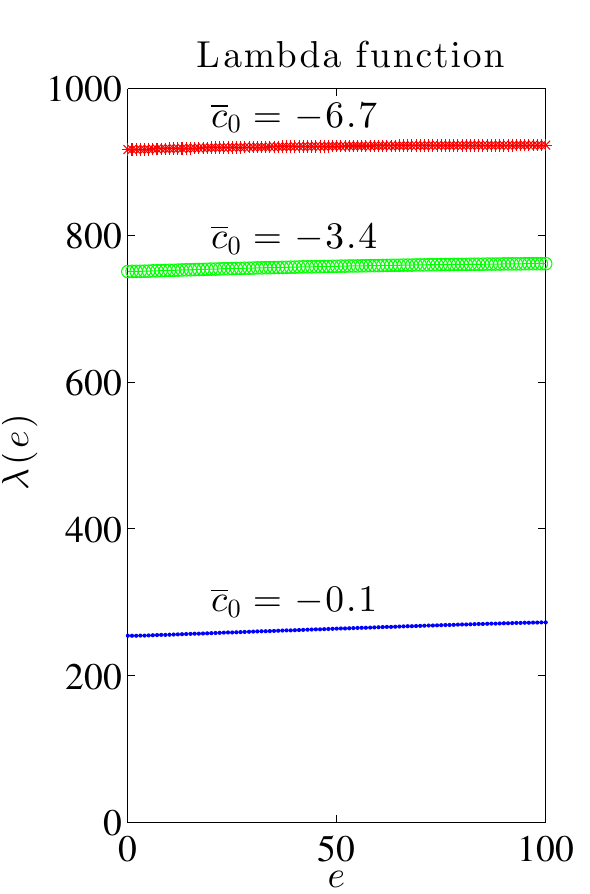}}
\caption[Non rechargeable node]{(a) Optimal thresholds for a harvesting node with $B=100$, stochastic energy costs, a unit-mean exponential importance distribution and $\gamma = 0.999$, for different values of {$\overline{c}_0$}. (b) The value function $\lambda(e)$.}
\label{fig:HarvestExp}
\end{figure}

Fig. \ref{fig:HarvestExp}.(a) shows the threshold function for each value of {$\overline{c}_0$}. {Note that, except for very small values of $e$,  threshold $\mu(e)$ is a decreasing function of $e$ (for very small values of $e$, the influence of $W(e)$ in \eqref{Eq.a_kOptStat2} is non-negligible). This can be explained as follows: for small values of $e$, the node increases the threshold to avoid that messages with low importance deplete batteries. For $e$ close to the maximum battery load, there is almost no benefit of refusing transmission, because the battery cannot be indefinitely charged and, thus, only very unimportant messages are censored.} Additionally, as {$\overline{c}_0$} gets smaller (more negative), $\mu$ gets smaller too. The reason is that faster battery recharge allows for a higher transmission rate.

\begin{figure}[t]
\centering
\includegraphics[trim=15mm 60mm 20mm 67mm, clip,width=0.8\columnwidth]{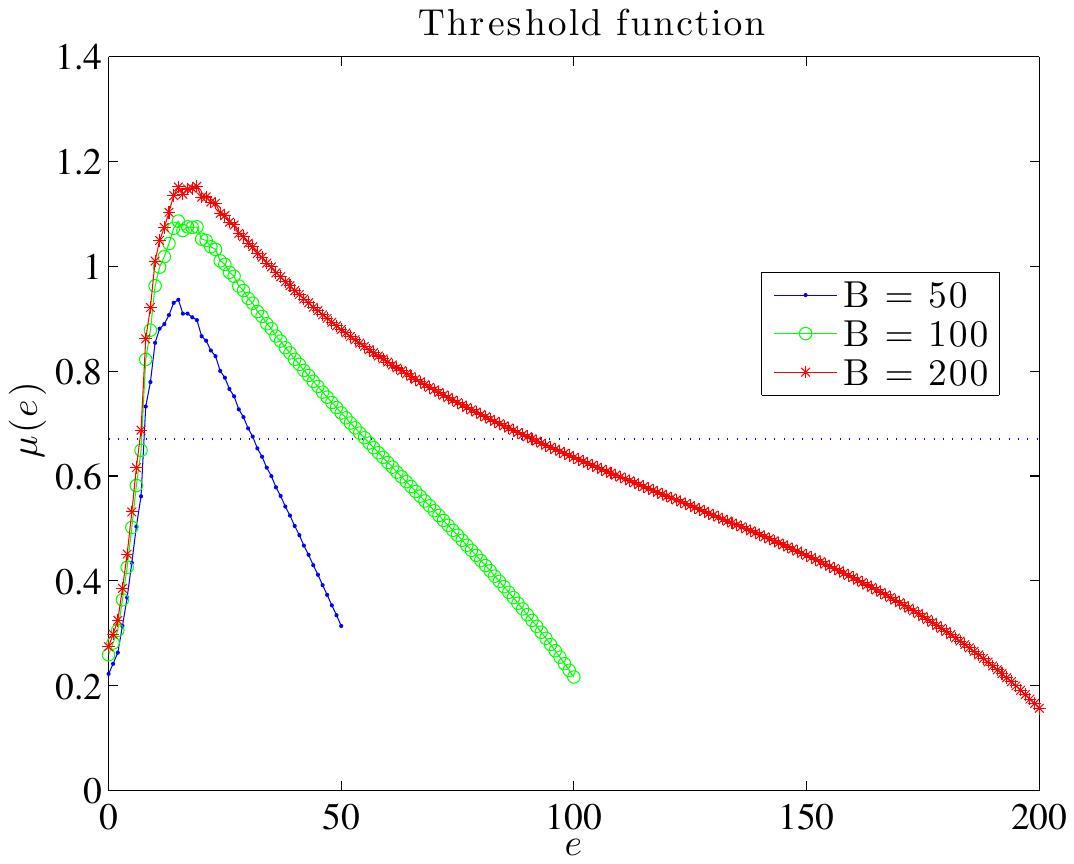}
\caption[Non rechargeable node]{Optimal thresholds for a harvesting node with ${\overline{c}_0}=-3.4$, ${\overline{c}_T=4}$, exponential importance distribution and $\gamma = 0.999$, for different values of the battery size. The horizontal dotted line shows the constant threshold value balancing the average energy consumption with the recharging rate.}
\label{fig:HarvestExp2}
\end{figure}

Fig. \ref{fig:HarvestExp2} illustrates the effect of changes in the battery size, $B$, for {the second test-case ($m_b=10$ and $\overline{c}_0=-3.4$}). We also show as a baseline a constant threshold policy $\bar{\mu}$ (horizontal dotted line), whose value is chosen so that the \emph{average} (long-term) energy consumption coincides with the \emph{average} harvested energy, without considering the battery limits. {This simple policy, which is related to the ``energy neutral operation'' concept proposed in \cite{Kansal2007}, is analyzed in different works, either under the name of balanced policy (BP) in \cite{michelusi2012optimal} or under the name of non-adaptive balanced policy (NABP) in \cite{michelusi2013transmission,gunduz2014designing}. Using this approach, the problem reduces to estimate the \emph{constant} threshold $\bar{\mu}$ that renders {$\EE\{c \} = 0$}, assuming $B=\infty$.} This is solved for
\begin{equation}
\bar{\mu} = F_X^{-1}\left(\frac{{\overline{c}_0}}
            {{\overline{c}_1}-{\overline{c}_0}}\right),
\label{EqMuBal}
\end{equation}
where {$\overline{c}_1 = \EE\{c |a=1\}$ and} $F_X^{-1}$ is the inverse of the cumulative distribution function of $x$.

\subsection{Steady-state distributions}
\label{sec:steady}

In this section, we will focus on asymptotic behavior, so that the effects of the initialization are disregarded.

Using elementary Markov Chain properties, it can be shown that, under some general conditions, the statistical distribution of the battery level converges, as $k$ goes to infinity, to a distribution $\boldsymbol{\phi}$ that is the solution of $({\bf I}-{\bf P})\boldsymbol{\phi} = {\bf 0}$ subject to {$\phi_i\geq0$ and} $\sum_{i=0}^{B} \phi_i=1$, where ${\bf P}$ be the transition probability matrix with entries
\begin{equation}\label{Eq.trans_prob}
p_{ij} = P\{e_k =j|e_{k-1}=i\}, \qquad i,j=0,\ldots, B
\end{equation}
and where, for notational convenience, we started the matrix indexing at 0.

Using the stationary distributions, the expected performance of a selective transmitter can be computed as
\begin{align}
V^* = \lim_{k\rightarrow\infty} \sum_{t=k}^{\infty} \gamma^{t-k} \EE\{a_t w_t x_t \}.
\label{EqLambdaGen}
\end{align}
Leveraging AS3, and using  \eqref{Eq.opt.a_k.Stat}, \eqref{EqErk1} can be substituted into \eqref{EqLambdaGen} to yield,
\begin{align}
{V^*  = \frac{1}{1-\gamma} \EE\{a_t w_t x_t \}
     = \frac{1}{1-\gamma} \sum_{e=0}^B \EE\left\{a_t w_t x_t |e_t=e \right\} \phi_e
        \label{Eq.Vopt0}}
\end{align}

{Taking into account that $w_t$ does not depend on $x_t$, $a_t$ and $e$, the expectation in the sum can be computed as
\begin{align}
\EE\{a_t & w_t x_t |e_t = e \}  \nonumber\\
     =& \EE\left\{ w_t x_t | a_t=1, e_t=e\right\} P\{a_t=1|e_t=e\}      \nonumber\\
     =& \EE\left\{ w_t | a_t=1, e_t=e \right\}
        \EE\left\{x_t | a_t=1, e_t=e\right\}    \nonumber\\
     &  \cdot P\{a_t=1|e_t=e\}      \nonumber\\
     =& W(e)
        \EE\left\{x_t | a_t=1, e_t=e \right\}
        P\{a_t=1|e_t=e\}      \nonumber\\
     =& W(e)
        \EE\left\{x_t u\left(x-\frac{\mu(e)}{W(e)}\right) \right\}
        \label{Eq.Vopt1}
\end{align}
Defining function $g$ as
\begin{equation}
g(\mu) = \EE\{u(x-\mu)x \} = h(\mu) + \mu (1-F_X(\mu))
\label{Eq.Defg}
\end{equation}
and substituting \eqref{Eq.Vopt1} into \eqref{Eq.Vopt0} we arrive at
\begin{align}
V^*  &= \frac{1}{1-\gamma} \sum_{e=0}^B g\left(\frac{\mu(e)}{W(e)}\right) W(e) \phi_e
   \label{Eq.Vopt}
\end{align}}

To compute $\phi$, one has to compute first the transition matrix ${\bf P}$, which depends on the censoring policy [cf. \eqref{Eq.trans_prob}]. This is accomplished in Appendix B.

Strictly speaking, optimizing \eqref{Eq.Vopt0} is only equivalent to optimizing \eqref{EqSOrder2} when $\gamma \rightarrow 1$. Nevertheless, \eqref{Eq.Vopt0} is useful to evaluate the long-term performance of our strategy.

Additionally, we compute the expected performance of the balanced policy \eqref{EqMuBal} based on a constant threshold $\bar{\mu}$. For such constant threshold, $\mu(e)=\bar{\mu} W(e)$, and \eqref{Eq.Vopt} can be simplified as
\begin{align}
V_{\bar{\mu}}  = \frac{g(\bar{\mu})}{1-\gamma}  \sum_{e=0}^B W(e) \phi_{\bar{\mu},e};
\label{Eq.Voptmu}
\end{align}
Finally, for a non-selective transmitter, $\bar{\mu}=0$, $g(\bar{\mu})=\EE\{x\}$ and
\begin{align}
V_0 = \frac{\EE\{x\}}{1-\gamma} \sum_{e=0}^B W(e) \phi_{0,e}.
\label{Eq.Voptmu0}
\end{align}

%

\begin{figure}[t]
\centering
\includegraphics[trim=10mm 65mm 10mm 70mm, clip,width=.96\columnwidth]{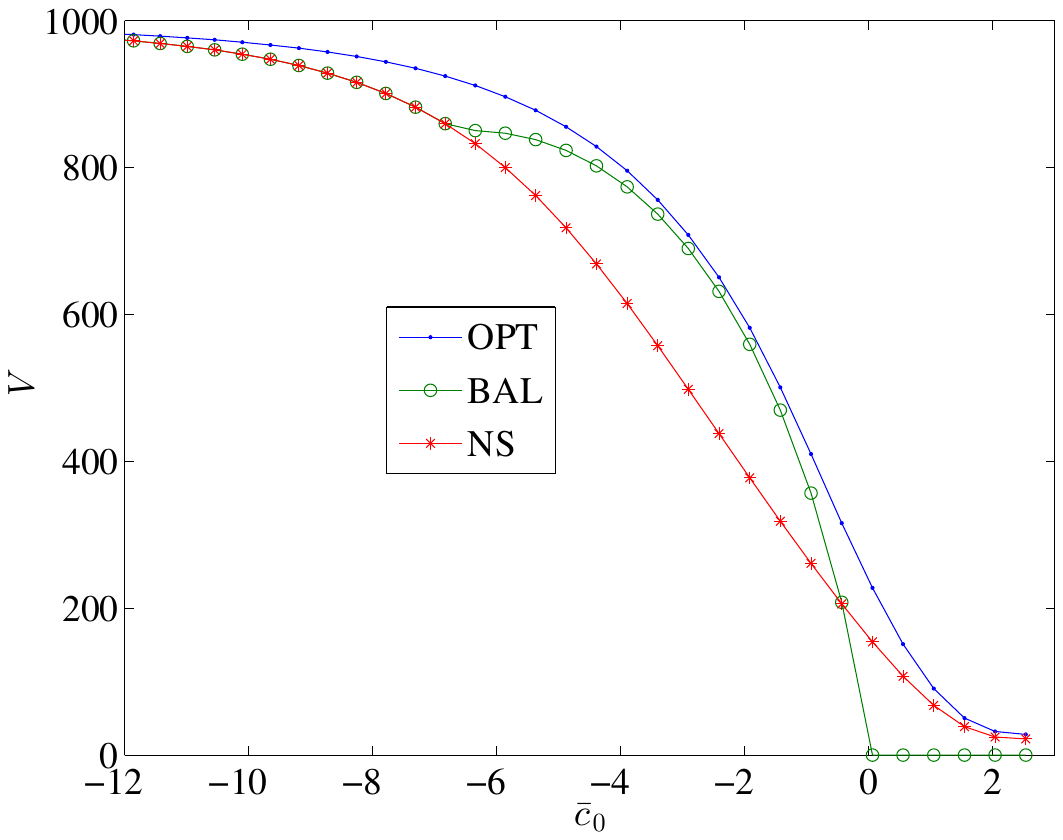}
\caption[Expected Performance]{Expected performance for a scenario with stochastic energy costs and high refill probability $p_b =1/3$, as a function of ${\overline{c}_0}$. Battery size $B=100$ and exponential distribution with unit mean and $\gamma = 0.999$.}
\label{Fig04HExp2StochStatLamb}
\end{figure}
\begin{figure}[t]
\centering
\includegraphics[trim=10mm 65mm 10mm 70mm, clip,width=.99\columnwidth]{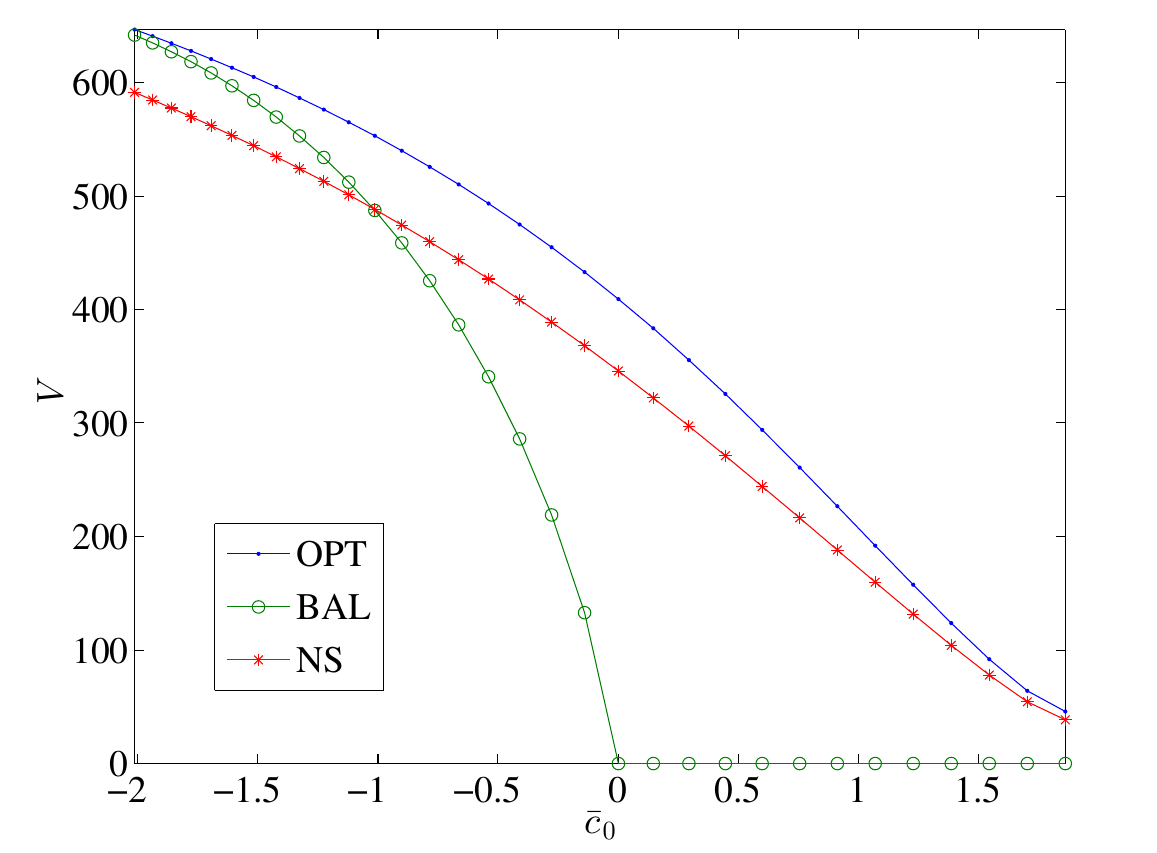}
\caption[Expected Performance]{Expected performance for a scenario with stochastic energy costs and low refill probability $p_b=0.04$, as a function of ${\overline{c}_0}$. Battery size $B=100$ and exponential distribution with unit mean and $\gamma = 0.999$.}
\label{Fig05HExpStochStatLamb}
\end{figure}

{Fig. \ref{Fig04HExp2StochStatLamb} shows the expected discounted reward [cf. \eqref{EqLambdaGen}] for $m_b$ ranging from 1 to 29  and for $p_b=1/3$. Fig. \ref{Fig05HExpStochStatLamb} illustrates a similar behaviour for a scenario with $c_T=c_R=2$ and $p_b=0.04$ respectively.} In the horizontal axis we show the corresponding value of $\overline{c}_0$ (average energy consumption when there is no transmission). {The figures demonstrate that: as $\overline{c}_0$ increases, the performance of the balanced transmitter (BAL) deteriorates much faster than the one of the optimal transmitter (OPT), being even worse than that a non-selective strategy (NS). This effect is very noticeable in Fig. \ref{Fig05HExpStochStatLamb}, where the harvesting is more occasional (lower probability of refill). Another relevant behavior is observed on the left region of Fig. \ref{Fig04HExp2StochStatLamb} (very negative values of $\overline{c}_0$). In that region, the energy harvesting is enough to compensate on average the communications costs, so that OPT, BAL and NS obtain the same performance. As energy decreases censoring is needed, but still OPT and BAL perform closely. Fig. \ref{Fig05HExpStochStatLamb} also shows that, even in situations where the transmissions costs are similar to reception costs, the performance of OPT is noticeably superior to BAL and NS. From both figures we can conclude that there are some scenarios where the performance of BAL is far from optimal.}

%

\section{Stochastic approximate schemes}
\label{S:Appr_sche}

The analysis in the previous sections provided insights on the behavior of optimal harvesting sensor nodes. However, the optimal policies presented so far are computationally very expensive, so that they can not be easily implemented in real time by sensors with limited computational capabilities. In this section, we present different ways to develop suboptimal adaptive stochastic schemes that reduce the computational complexity and, additionally, are able to deal with non-stationarities.


\subsection{A stochastic approximation to the optimal policy}\label{S:adapt_MDP}

The {\emph{threshold-based}} optimal policy presented in Section \ref{S:Opt_sel_fw} stands on two main assumptions on the energy dynamics: (a) {neither the energy consumption nor the recharge depend on the importance value}, but only on the taken action, and (b) in the linear regime (i.e., with the exception of the battery saturation points) the variation of the energy stored in the battery does not depend on the current battery level.

The main difficulty {to obtain the optimal solution} using the {value iteration} method proposed in \eqref{Eq.Mu_1c} and \eqref{Eq.Lambda1c} is the computation of the expectations involved. { But if we assume again that $e_k$ is discrete}\footnote{{Even if not discrete, a common strategy to deal with the estimation of continuous (state) policies is to discretize the input state and use linear interpolation.}}, they can be stochastically approximated in a sample-based manner.

{In order to do so, we will represent the policy as an instance of Robbins-Monro algorithm \cite{yin2003stochastic} and use stochastic approximation techniques to get our algorithm. First of all, we will decompose cost $c_k$ as
\begin{equation}
c_k = c_{0,k} + a_k \ctk
\end{equation}
that is, $c_{0,k}$ represents the energy consumption when a message is censored ($a_k=0$) and $\ctk$ represents the incremental cost of transmitting a message. In addition, it is useful to write \eqref{Eq.Mu_1c} and \eqref{Eq.Lambda1c} in matrix form.} Let us define vectors $\boldsymbol{\lambda} = (\lambda(0),\lambda(1),\ldots,\lambda(B))^\intercal$, and $\boldsymbol{\omega} = (W(0),\ldots,W(B))^\intercal$, and the vector of success indices ${\bf w}_{c} = (u(0-c), u(1-c), \ldots, u(B-c))^\intercal$. We assume in the following that vectors are indexed from 0, in such a way that, for instance, $\lambda_e = \lambda(e)$. Also, we define the transformation $\boldsymbol{\lambda}' = {\bf T}_c \boldsymbol{\lambda}$ such that $\lambda'_e = \lambda_{\clip_B(e-c)}$ . In Appendix C we derive the following adaptive rules as an instance of Robbins-Monro algorithm \cite{yin2003stochastic}.
\begin{align}
\label{AlgOmega}
\boldsymbol{\omega}_{k+1}
    &= (1-\eta_k) \boldsymbol{\omega}_k + \eta_k {\bf w}_{{c_{0,k}+\ctk}} \\
\label{AlgAlpha}
\boldsymbol{\alpha}_{k+1}
    &= (1-\eta_k) \boldsymbol{\alpha}_k
     + \eta_k {\bf T}_{{\bf c}_{0,k}} \boldsymbol{\lambda}_k \\
\label{AlgBeta}
\boldsymbol{\beta}_{k+1}
    &= (1-\eta_k) \boldsymbol{\beta}_k
     + \eta_k {\bf T}_{{{\bf c}_{0,k} + \ctk}} \boldsymbol{\lambda}_k  \\
\label{AlgLambda}
\boldsymbol{\lambda}_{k+1}
    &= (1-\eta_k)\boldsymbol{\lambda}_k + \eta_k
       \left(\gamma \boldsymbol{\alpha}_k
           + \left(x_k \boldsymbol{\omega}_k
           - \gamma \left(\boldsymbol{\alpha}_k-\boldsymbol{\beta}_k \right) \right)^+\right)
\end{align}
%
where $\eta_k$ stands for the learning stepsize, which can be {set either} to diminish with time (for instance in stationary scenarios where one wants $\mu_k$ to converge to a fixed function) or to a small constant (for adaptation to changes in non-stationary scenarios). {The stepsize} must be chosen in order to balance a good convergence speed and a low steady-state error, but always satisfying the Robbins-Monro {stepsize conditions} for convergence \cite{yin2003stochastic}.


Note that, at each iteration, the above rules update all components of vector $\boldsymbol{\lambda}$, i.e., the whole estimate of $\lambda(e)$ is updated for all values of $e$. Thus, the computational load and memory requirements grow linearly with the number of discrete energy values. {Neither} the importance nor the energy distribution are required to be {known to use \eqref{AlgOmega}, \eqref{AlgAlpha} and \eqref{AlgBeta}. Instead, they are run every time a sample of} $c_{0,k}$ or {$\ctk$} is observed. Regarding the observability of $c_{0,k}$ and {$\ctk$} some remarks {are in order}.
{\begin{itemize}
\item When $a_k=0$, $\ctk$ is not observed, and $\boldsymbol{\beta}_k$ is not updated.
\item When $a_k=1$, we assume that the sensor can measure the energy level right before taking the necessary actions to transmit the packet. Thus, the node can measure an intermediate energy level $e_{k'}=\phi_B(e_k-c_{0,k})$, that can be used to estimate $c_{0,k}$ and $\ctk$, as follows:
\begin{itemize}
\item If there is no battery underflow or overflow, then $c_{0,k}= e_k-e_{k'}$ and $\ctk=e_{k+1}-e_{k'}$.
\item If there is battery depletion, and $e_{k+1}=0$, $\boldsymbol{\alpha}_k$, $\boldsymbol{\beta}_k$ and $\boldsymbol{\omega}_k$ are not updated.
\item If there is battery overflow ($e_{k+1}=B$), we estimate $c_{0,k}$ and $\ctk$ as $\tilde{c}_{0,k}= e_k-e_{k'}$ and $\tildectk=e_{k+1}-e_{k'}$.
\end{itemize}
\end{itemize}
Note that the cost estimates under battery overflow are biased, and more involved \emph{imputation methods}, see e.g. \cite{ghahramani1994supervised}, could be applied to solve this issue. They are left for future work.} In Table \ref{fig:SAP}, we summarize the main steps required to run the stochastic approximate policy (SAP) algorithm.

\begin{table}[ht]
	\centering
	\begin{normalsize}	
	\begin{tabular} {l}
		\hline
		\textbf{SAP algorithm} \\
		INPUTS: Initial battery $e_0$, $\gamma$, and $\eta$	\\
		\hline\vspace{-.2cm}
		\\
		Initialize $\boldsymbol{\omega}_0={\bf 0}$, $\boldsymbol{\lambda}_0={\bf 0}$,$\boldsymbol{\alpha}_0={\bf 0}$,$\boldsymbol{\beta}_0={\bf 0}$. \\
        \\		
		At each time step $k$, at the sensor node:\\							
		\ 1. Sense or Receive a message of importance $x_k$. \\
		\ 2. Harvest energy $b_k$ and consume $\hat{c}_{0,k}$, (${c}_{0,k}=\hat{c}_{0,k}-b_k$) \\
		\ \ \ \ \  $\bullet$ Energy after sensing $e_{k'} =\phi_B( e_{k} - {c}_{0,k})$.\\	
		\ 3. Decide about transmitting the message: \\
		\ \ \ \ \ $a_k = u(\omega_{e_{k}} \cdot x_k - \mu_{e_{k}})$.\\
		\ 4. Consume additional cost $\ctk$ if $a_k=1$,\\
		\ \ \ \ \ $\bullet$ {$c_k = c_{0,k}+ a_k \ctk$}\\
		\ \ \ \ \ $\bullet$ $e_{k+1} = \phi_B({e_{k} - c_k}) $.\\
		\ 5. Approximate $c_{0,k}$ as $\tilde{c}_{0,k}=e_{k'}-e_{k}$ \\
		\ \ \ \ \  and $\ctk$ as  $\tildectk = e_{k+1}-e_k'$.  	
		\\
		\ 6. Update $\boldsymbol{\lambda}_k$ according to \eqref{AlgLambda}. \\
		\ 7. If ($e_{k+1} > 0$): {update $\boldsymbol{\alpha}_k$ using \eqref{AlgAlpha} with} $\tilde{c}_{0,k}$. \\
		\ 8. If ($e_{k+1} > 0$ and $a_k=1$): update $\boldsymbol{\omega_k}$ and $\boldsymbol{\beta}_k$\\
		\ \ \ \ \ according to \eqref{AlgOmega} {and} \eqref{AlgBeta} using $\tilde{c}_{1,k}$. \\		
		\ 9. $\boldsymbol{\mu}_k = \gamma ( \boldsymbol{\alpha}_k-\boldsymbol{\beta}_k)$. \\
		\hline			
	\end{tabular}
	\end{normalsize}
	\caption{Description of stochastic approximate policy (SAP) algorithm.}
	\label{fig:SAP}	
\end{table}


\subsection{Q-learning}\label{S:q_learn}

An alternative design is to use universal stochastic approximation methods that do not require any assumption on the state dynamics, like $Q$-learning \cite{sutton1998reinforcement}. They can be expected to outperform SAP in scenarios where the above assumptions are too unrealistic. However, there is a price to pay for this flexibility. The SAP algorithm {leverages} the structure of the optimal decision to reduce the search space and speed up convergence{.  On the other hand,} $Q$-learning {has} to compute a $Q$ value for each possible action and state (energy and importance value); as a consequence, the memory requirements may be too high. Furthermore, at each iteration, the {$Q$-learning} algorithm only updates the estimate of the value functions \emph{at the current state}, and though convergence can be theoretically guaranteed, it requires to visit all possible states infinitely often \cite{wiering2012reinforcement}. In practice, for large state spaces, convergence to the optimal solution is difficult and learning time is much {larger} than that of model-based approaches as it will be shown in the numerical experiments in Section \ref{S:Simulations}.

Our $Q$-learning implementation is a minor variation of the algorithm proposed in \cite[Eq. (8)]{Blasco2013} for a similar application. In order to be able to apply it to our setup, we quantized the importance value, which is a real number, into a number of levels (standard $Q$-learning {needs} a discrete state space), and apply the algorithm in \cite{Blasco2013}. This algorithm has two free parameters the learning rate $\alpha_k$, and $\epsilon$ the exploration probability in the \emph{$\epsilon$-greedy action selection} method.

\subsection{Adaptive balanced transmitter}
\label{S:adapt_balanced}


A further step to decrease computational complexity is to restrict the attention to suboptimal policies that are easy to compute. A good candidate is the balanced policy presented in \eqref{EqMuBal}. This policy {estimates} a constant (\emph{energy independent}) threshold that {tries} to balance {(on average)} the harvested and consumed energy under the infinite battery assumption.

The main difficulty of solving \eqref{EqMuBal} is that it requires knowledge of the importance distribution. In most cases such a knowledge is not available, or the distribution may not be stationary. In the next lines, we present a adaptive scheme to bypass those problems. An equivalent method for obtaining an adaptive balance policy was presented in \cite{fernandez2013_DP_and_DualStoch}.

Upon defining $\rho =\frac{\overline{ c}_1   }{\overline{ c}_1  -\overline{ c}_0  }$, \eqref{EqMuBal} states that the constant threshold $\bar{\mu}$ is the $\rho$-quantile of the distribution function of $x$, $F_X$. Based on the results in \cite{saerens2000building}, which builds functions that are minimized at specified statistics, we can define the following cost function whose expectation attains its minimum at the $\rho$-quantile of some conditional probability density
\begin{equation}
J(x, \bar{\mu}) = \rho·(x-\bar{\mu})^+ + (1-\rho)(\bar{\mu} - x)^+.
\label{Eq.xentropy}
\end{equation}
To minimize \eqref{Eq.xentropy} we implement a stochastic gradient method
\begin{equation}
\mu_k = \mu_{k-1} + \eta_k \big( \rho_k u( x_k - \mu_{k-1} )
      - (1-\rho_k)  u(\mu_{k-1} - x_k ) \big),
\label{Eq.adaptive_balanced}
\end{equation}
where $\eta_k$ represents, again, a learning step. Both $\overline{ c}_1 $ and $\overline{ c}_0 $ have also to be estimated in order to calculate $\rho$. This can be easily accomplished using the sample mean.

{In the following, this method will be referred to as \emph{Adaptive Balanced Transmitter} (ABT)}.

\section{Numerical experiments}\label{S:Simulations}

In this section, we run simulations to compare the performance of the three presented stochastic policies: SAP, described in Table \ref{fig:SAP}{; $Q$-learning;} and {ABT}, given by \eqref{Eq.adaptive_balanced}. A non-selective scheme (NS) is also included as a baseline and, when possible, the theoretical optimal performance (OPT), calculated as in Section \ref{Sec.SingleNode}. Three sets of {numerical experiments are simulated}. The first one {analyzes} a single-hop network with stationary energy harvesting processes. The second one {considers} a non-stationary energy harvesting {scenario,} where the statistics of the energy refill process vary in a periodical way. {The third one analyzes the behavior of the developed schemes in multi-hop networks. Although the algorithms proposed in this paper are not optimal for that case, the idea is to test their performance and compare it with other existing alternatives.}

To run the experiments we assume that: 1) Nodes have a way to measure the energy consumed after each decision. 2) Performance is measured in terms of a sample-based estimation ($\hat{V}$) of the steady-state discounted aggregate reward received at the destination, and, when possible, it is compared to the optimal {discounted} value obtained using (\ref{Eq.Vopt}). $\hat{V}$ is calculated as the discounted ($\gamma$-weighted) sum of the importance values of the messages received at the sink {during the second half of the simulated horizon}. 3) In all experiments, the value of the reward discount factor $\gamma$ is set to 0.999. 4) Message importances follow an exponential distribution (with mean {$\overline{x} = 2$}), which is a sensible approximation for several practical scenarios, such as monitoring {applications,} where most messages are of low importance and a small number of them (alarms) have a high importance. For the non-harvesting case, \cite{ArroyoVallesEtAl09} found that these assumptions on the importance distribution were not very critical, and the main conclusions there can be extended to this paper too. {In any case,} neither {ABT} nor SAP nor $Q$-learning require prior information about the importance/energy distributions.

\subsection{Single-hop network}

\subsubsection{Stationary energy refill}

The first set of examples is aimed at validating the {ABT} and SAP schemes. {{The idea is to show} that they are a good approximation to the optimal decision {rule}. {In this section,} a single-hop (cellular) network is considered. {This allows us to assume} that if the energy available in the battery is greater than that required to transmit a message, the message is successfully received at the \emph{sink}. For this experiment, the learning rate of both adaptive algorithms is set to $\eta_k = 1 /( 1 + \delta \cdot k )$, with $\delta<1$ denoting a small constant. Since all processes are stationary, convergence is then guaranteed. {For each of the schemes, the value of $\delta$ is selected to maximize the value of $\hat{V}$. For the comparison with $Q$-learning, we quantized the importance value into 100 different levels, used a learning rate $\alpha=0.2$ and random exploration with probability $\epsilon = 0.1$ \cite{powell_book}. The energy refill values are drawn from a Bernoulli distribution such that the harvested energy at each slot (if non-zero) is $e_H=30$ with a fixed probability of harvesting (the value of the probability will vary to simulate different harvesting scenarios). Moreover, we use the cost model in \eqref{eq:cost_model0} and \eqref{eq:cost_model1}, with $c_R=3$, $c_T = 5$, and assuming that the message is retransmitted until it is correctly received (the error probability is set to $0.3$}). As already pointed out, by varying the value of the probability of refill, we vary the value of ${\overline{c}_0}$ and ${\overline{c}_1}$. The results presented next are obtained after averaging 100 simulations with different random message sequences and energy patterns.}

\begin{figure}[htb]
  \centering
  \includegraphics[trim=10mm 120mm 10mm 110mm, clip,width=1\columnwidth]{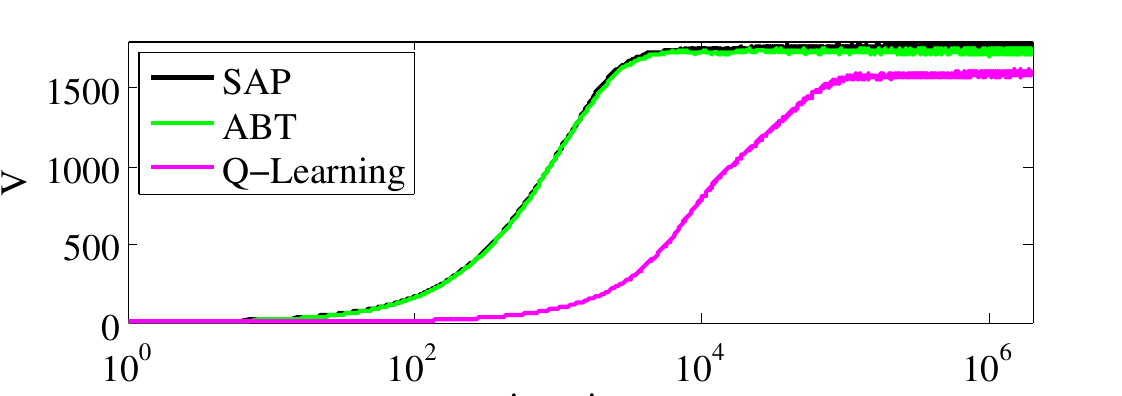}
\caption{Comparison between the SAP, {ABT}, and $Q$-learning algorithms with learning rate $\alpha=0.2$ and random exploration with probability $\epsilon =0.1$.}
\label{fig:Qlearning}
\end{figure}

Fig. \ref{fig:Qlearning} (where the probability of harvesting is $0.3$, and therefore ${\overline{c}_0}=-6$) confirms the expected behavior: i) {the convergence speed of $Q$-learning (non parametric, non model-based)} is much slower than that of SAP and {ABT}; and ii) within the simulated time horizon {(note the logarithmic scale in the horizontal axis),} $Q$-learning does not converge to the optimal solution, mainly because {it} does not visit some of the states (hence, the corresponding values of the {action-value} function can never be properly estimated). {Due to these limitations, $Q$-learning will not be included in any other simulation.} It is important to remark that the use of more complex reinforcement learning algorithms, e.g.{,} based on function approximation schemes, is likely to perform well in this {class} of problems. We consider this an interesting research direction {to be explored} in future works.

\begin{figure}[ht]
\centering
\includegraphics[trim=10mm 65mm 10mm 90mm, clip,width=1\columnwidth]{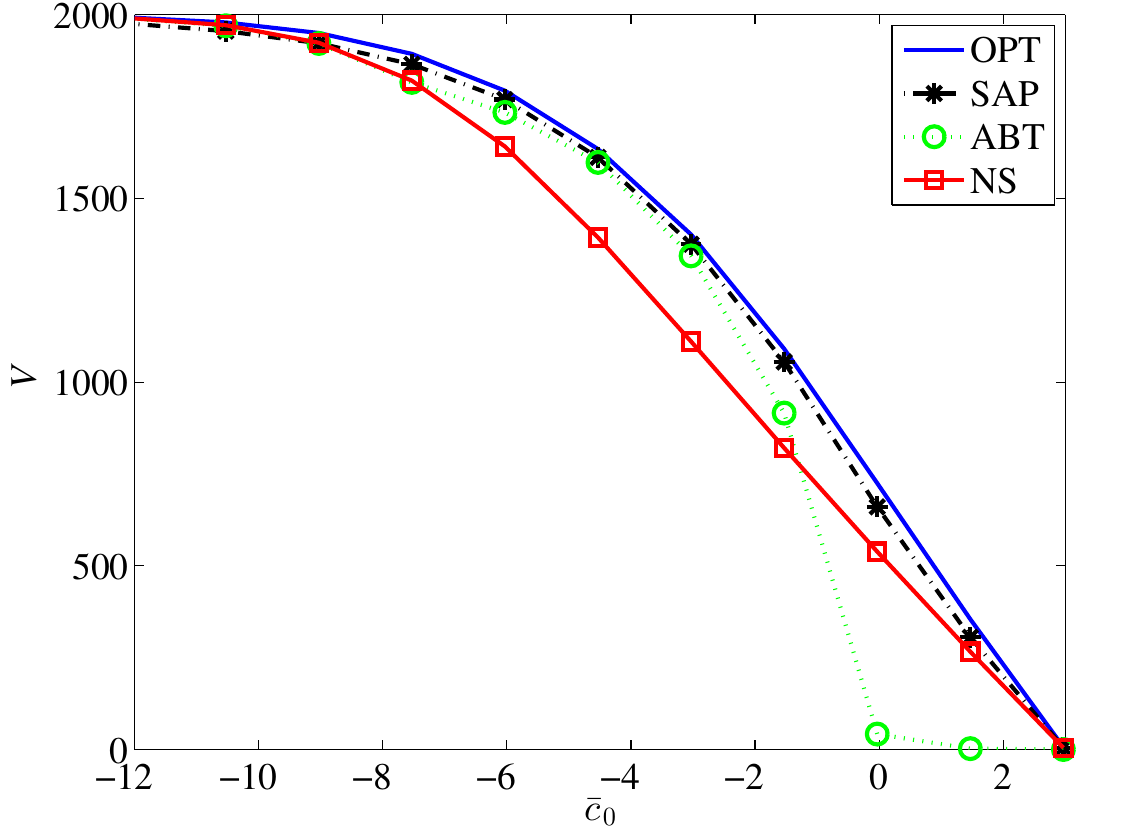}
\caption[]{Performance in terms of $\hat{V}$ of the proposed algorithm compared to the theory for different ${\overline{c}_0}$ values for one node simulation.}
\label{FigExperimentSingle1}
\end{figure}

\begin{figure}[ht]
\centering
\includegraphics[trim=10mm 65mm 10mm 90mm,clip,width=1\columnwidth]{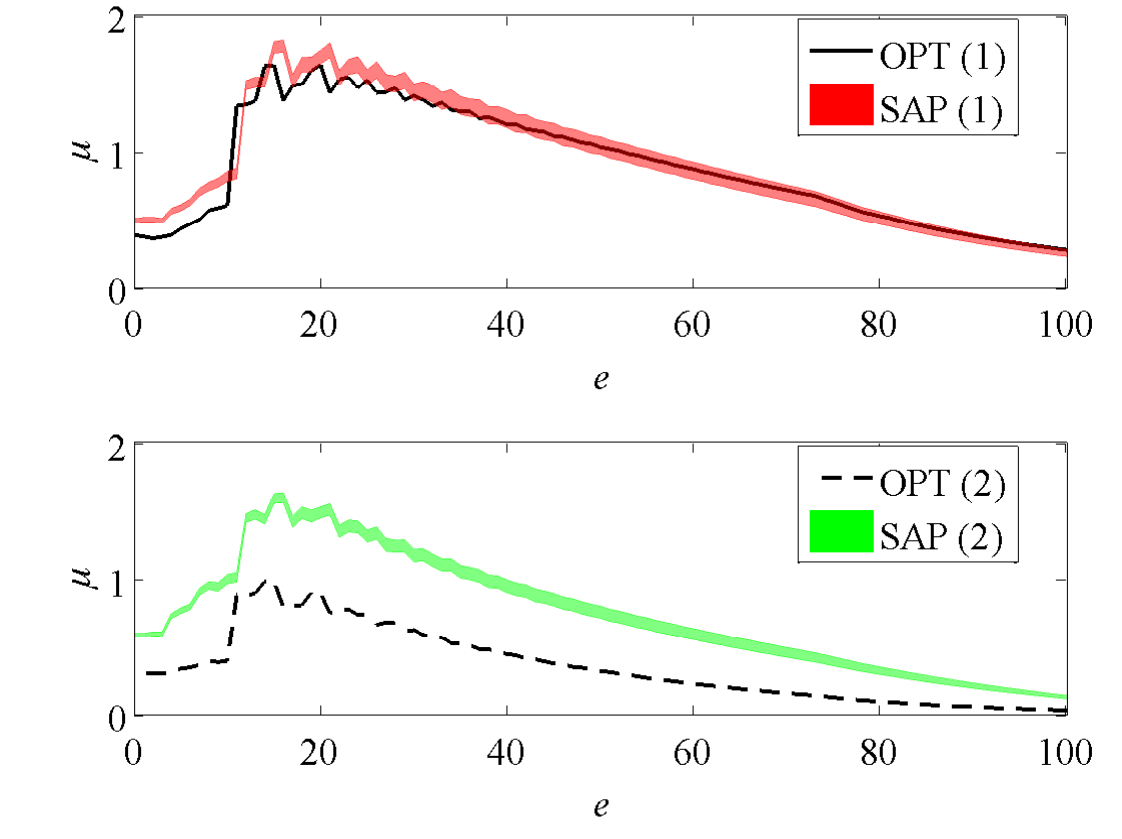}
\caption[]{Estimation of $\mu$ obtained by the SAP (shaded area) and OPT (solid and dashed line) schemes for (1) ${\overline{c}_0} = - 3$ and (2) ${\overline{c}_0} = - 9$. For the approximated $\mu$, the average value $\pm 2\sigma$ is shown.}
\label{FigExperimentSingle1Thres}
\end{figure}

In Fig. \ref{FigExperimentSingle1}, where the probability of harvesting varies from $0.001$ to $0.5$, we plot the average performance $\hat{V}$ versus ${\overline{c}_0}$ for {SAP, ABT and NS,} together with the optimal {performance} (OPT) calculated {using \eqref{Eq.Vopt}}. It is clear that the SAP scheme outperforms the {ABT} scheme for all tested cases. More importantly, the SAP scheme achieves a performance very similar to that of the OPT scheme. Fig. \ref{FigExperimentSingle1Thres} shows that the estimated thresholds are {also} close to the optimum ones. In case 2), where ${\overline{c}_0} = - 9$, the energy costs are typically underestimated (due to battery overflows), and consequently the estimation of the threshold is biased. Nevertheless, the performance of SAP is almost optimal {(cf. Fig. \ref{FigExperimentSingle1})}.

In a nutshell, the simulations validate our approach for the tested scenarios and demonstrate that the stochastic approximation is able to approximate the correct thresholds and, consequently, achieves an almost optimal performance.

\subsubsection{Non-stationary energy refill}

Table \ref{TabExperimentSingle2} shows that, although our derivations assumed that the processes were stationary, our proposed scheme can be applied to non-stationary scenarios. Specifically, we simulate a periodic refill for which ${\overline{c}_0}$ is positive during some periods of time (when the harvested energy does not compensate the operating cost) and negative {during others}. The actual refill in each slot follows a Bernoulli distribution with probability 0.3 and with different $e_H$ in the two regimes. {This way,} we simulate a simplified version of a number of harvesting devices that have a periodical behavior with some random component, such as solar energy harvesters.

{Since the} environment is not stationary, {the learning rate is set to a constant value that trades off convergence rate and (gradient) noise in ``steady state''}. In this experiment, we use $\eta=0.5$ for SAP and $\eta=0.05$ for  {ABT}, which {are, respectively,} the values that empirically maximize $\hat{V}$. The use of an heuristic rule for stepsize selection in a real application is an important issue, but it is out of the scope of this paper.

\begin{table}[ht]
\centering
\begin{tabular}{|c|c|c|c|}
\hline                  & SAP      & {ABT} & NS \\
\hline $\hat{V}$ (mean) &  1188.32 & 947.93 & 685.94 \\
\hline $\hat{V}$ (std)  &  3.25    & 2.39   &  3.23 \\
\hline
\end{tabular}
\caption{Mean and standard deviation of $\hat{V}$ achieved by the SAP, {ABT} and NS algorithms in a non-stationary environment. Listed values were obtained by running 200 different simulations.}
\label{TabExperimentSingle2}
\end{table}

As already mentioned, the results listed in Table \ref{TabExperimentSingle2} {show} that the benefits of our stochastic approximation schemes also hold true in {{the tested} non-stationary environment. The {time} evolution of the battery load with time in Fig. \ref{FigExperimentSingle2batt} provides additional insights. It shows that, when no censoring is applied, the battery of the node is empty {during} long periods of time, leading to a poor performance. {On the other hand,} the adaptive balanced transmitter ({ABT}) is able to keep some small amount of {energy in the battery} during the low-refill periods {and, hence, it achieves a better performance}. More importantly, the proposed stochastic approximation method is able to modify (adapt) rapidly enough the transmission threshold when the refill pattern changes, so that it always has some battery to transmit important messages and, hence, it obtains the best performance.

\begin{figure}[ht]
\centering
\includegraphics[trim=10mm 60mm 10mm 160mm, clip,width=1\columnwidth]{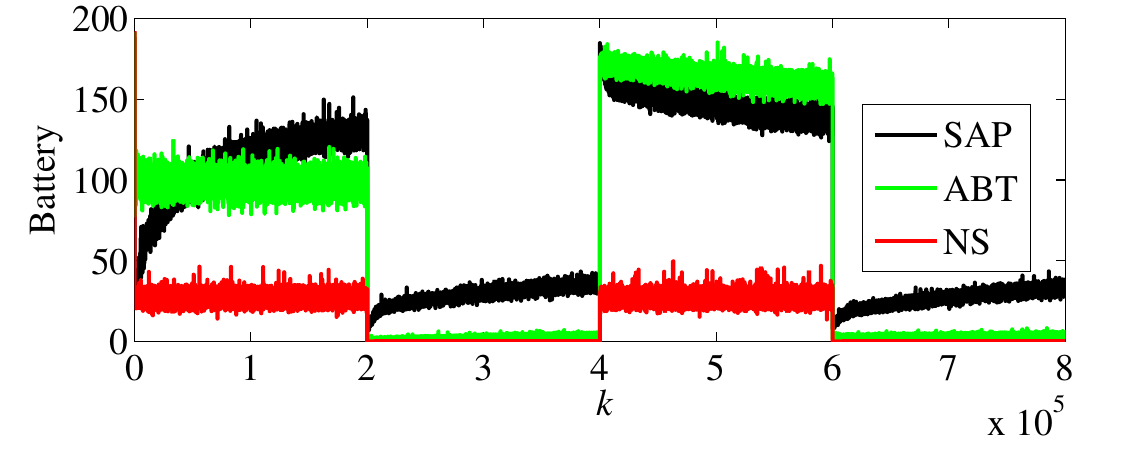}
\caption[]{Battery evolution (across time) of SAP, {ABT} and NS algorithms in a non-stationary environment.}
\label{FigExperimentSingle2batt}
\end{figure}

\subsection{{Multi-hop networks}}

{In this section, we test our algorithms in a more complex scenario: a multi-hop network with two different topologies. Fig. \ref{fig:net_topology}.(a) represents a random tree topology with a variable number of nodes, where all messages are routed through the tree to the root (sink) node, which is assumed to be wired to a power line. The network tree is randomly constructed so that for each node of index $i$, we choose with equal probability one and only one node of index $j$ (with $j > i$) to be connected with. The tree graph can be understood either as the actual topology of a cycle-free network or as the spanning tree obtained after running a specific routing protocol to a network with cycles. On the other hand, Fig. \ref{fig:net_topology}.(b) represents a squared-lattice network with fixed number of nodes. In this case, the routing to the sink is fixed and computed using the Dijkstra algorithm. Although the networks simulated are relatively simple, they are useful to illustrate the interaction between the censoring strategies across different nodes.}

{In addition, we} consider a success index $w_k=1$ when the node is able to transmit the message, consequently the scheme is suboptimal relative to the maximization of the discounted aggregated reward of the messages received at the sink. It is also assumed that all nodes generate messages with the same probability, so that nodes closer to the sink {will handle more traffic.} As in the previous subsection, the importance of the messages is an exponential i.i.d. process with mean 2.

\begin{figure}[htb]
  \centering
  \includegraphics[clip,width=0.7\columnwidth]{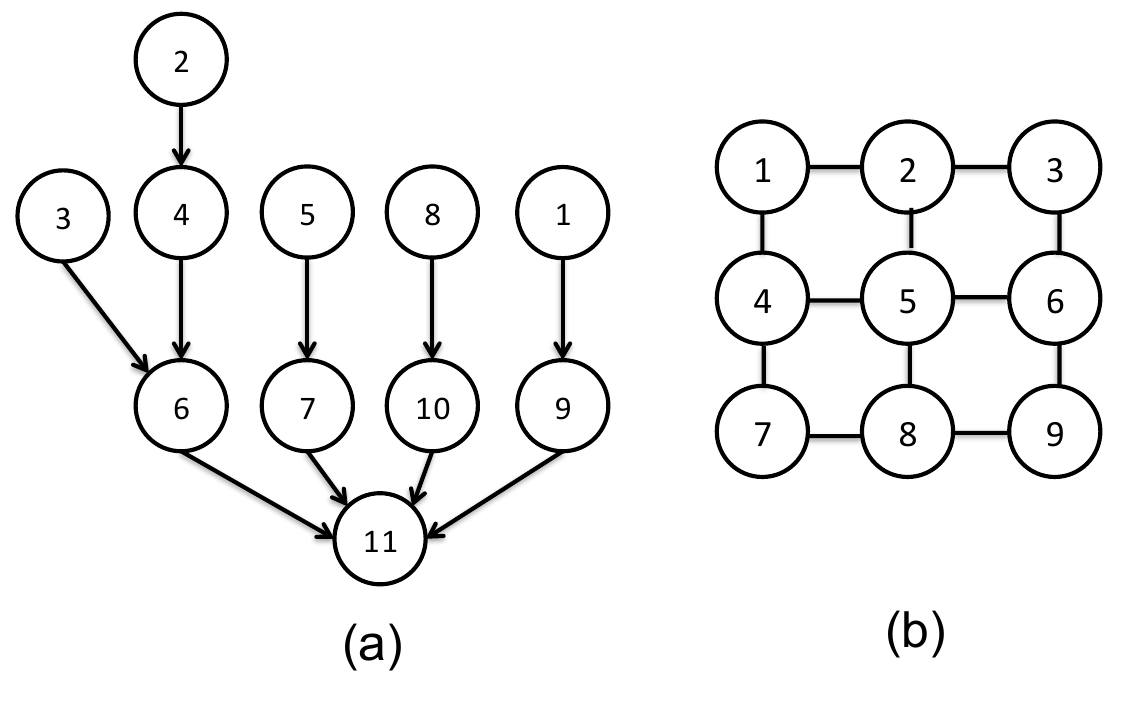}
\caption{Example of network routing topology. All messages are rooted to sink node 21, which is connected to a power line and has no battery limitations.}
\label{fig:net_topology}
\end{figure}

{We simulated 6 different scenarios, all of them with fixed $c_R$ and $c_T$, and without channel losses. The harvesting is random and different harvesting probabilities are considered.} Scenarios 1, 2, and 3 correspond to a tree network [Fig. \ref{fig:net_topology}.(a)] with 20 nodes (scenarios 1 and 2) and 10 nodes (scenario 3). The harvesting probability is $0.1$ for scenario 1, and $0.5$ for 2, 3. Scenarios 4, 5, 6 correspond to a grid network of 9 nodes (8 sensors and a sink). In scenarios 4 and 5, the sink is located at one of the corners, e.g., node 9 in Fig. \ref{fig:net_topology}.(b); while for scenario 6 it is at the center of the grid (node 5 in Fig. \ref{fig:net_topology}.(b)).} {The harvesting probability is $0.5$ for scenario 4 and $0.9$ for scenarios 5 and 6.} {Fig. \ref{FigExperimentNet} shows the average of 200 simulations.}


\begin{figure}[ht]
\centering
\includegraphics[trim=12mm 70mm 10mm 150mm, clip,width=1\columnwidth]{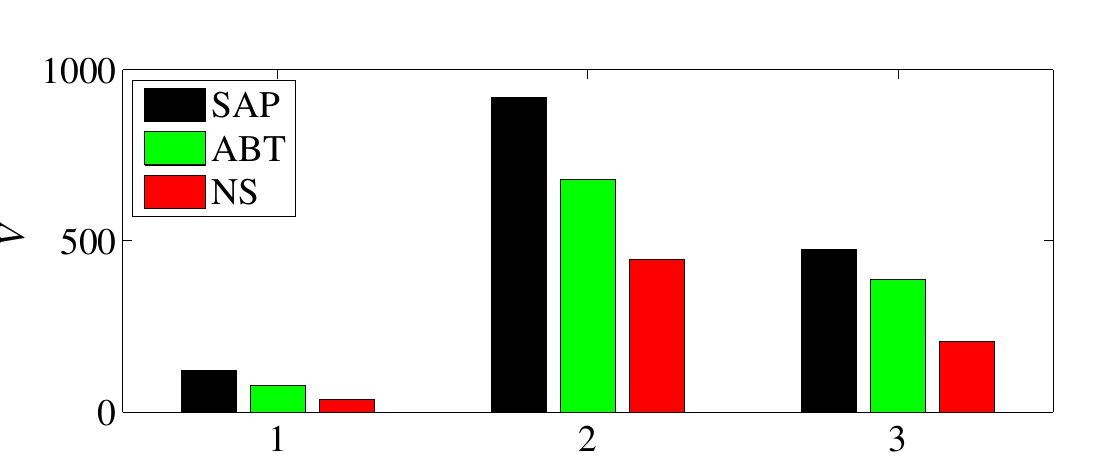}
\includegraphics[trim=12mm 60mm 10mm 150mm, clip,width=1\columnwidth]{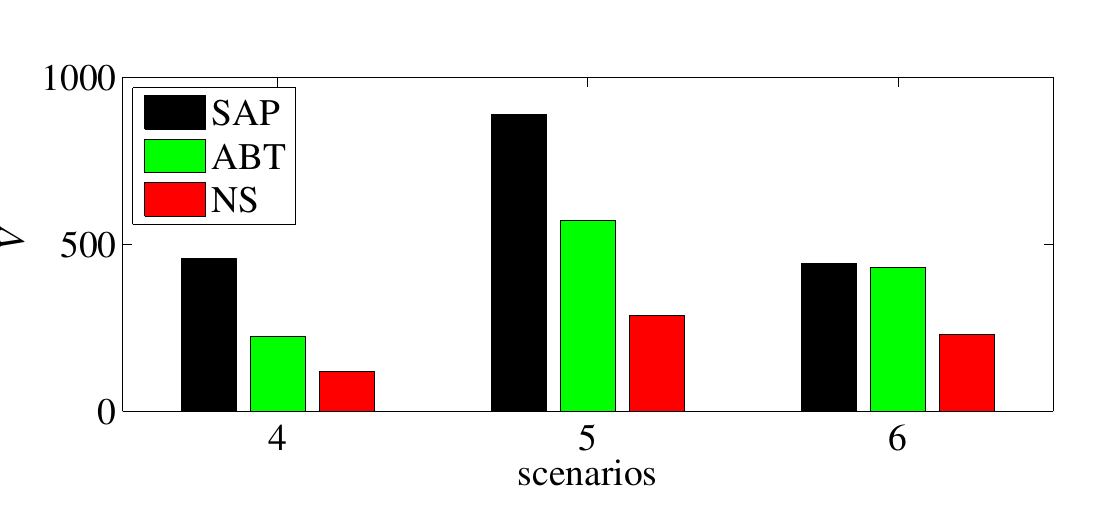}
\caption[]{Performance in terms of $\hat{V}$ of the two proposed algorithms and the non selective method for 6 different multihop scenarios.}
\label{FigExperimentNet}
\end{figure}

{The results point out that although the algorithms were not explicitly designed for multi-hop networks, they achieve a better performance than that of the tested alternatives. In fact, the gain of the SAP scheme relative to the ABT scheme in Fig. \ref{FigExperimentNet} is in scenarios 2,4,and 5} much larger than that observed in the experiments presented in the previous section. Hence, the combination of the local optimization processes at each node has a positive global influence. {Although the results are not comprehensive, they serve as a preliminary validation. Designing decisions jointly across nodes, accounting for the costs of exchanging information, or investigating the effect of interference and medium access control are all aspects worth analyzing (for example, \cite{michelusi2013optimal} showed that balanced policies are suboptimal if interference is present), but they are out of the scope of the present manuscript and are left as future work.}



\section{Conclusions}\label{S:Conclusions}

In this paper, we designed and analyzed a censoring scheme for WSNs with harvesting devices. The problem is modeled using the MDP framework and some assumptions are made in order to obtain a threshold-based optimal policy. Some insights about these optimal policies are provided and suboptimal schemes, based on stochastic approximation, are developed too. {Numerical experiments confirmed the theoretical claims, showed the benefits of our approach with respect to previous works, specially when the harvested energy is scarce. Finally, experiments showed that these schemes perform well even if some of the assumptions under which they were designed (i.e stationarity, single-hop networks) do not hold.}


Current and future research include accounting explicitly for multi-hop networks, designing alternative stochastic approximations schemes, {including more complex data and harvesting models, such as Markov and Generalized Markov Models \cite{ho2010markovian}}, rendering the schemes robust to imperfections in the state, and considering soft-decision schemes \cite{michelusi2012operation}.

\appendices

\section*{Appendix A: Proof of Theorem \ref{Th.General}}
\label{ProofThGeneral}

Using \eqref{EqReward}, and for ${\bf s} = (e,x)$,
\begin{align}
\EE\{r_k|a_k=a,{\bf s}_k={\bf s} \} = a x W(e,x)
\label{EqCondrk}
\end{align}
where $W(e,x) = \EE\{w_k|e_k=e,x_k=x \}$.

Also, using \eqref{Eq:EnergyDynamics} and taking into account that {$x_k$ is an i.i.d. sequence and independent {of} $e_k$ (as1), and $c_k$ is independent of $x_k$ and $e_k$ given $a_k$ (as2){, we have that}}
\begin{align}
\EE\{V_{\pi^*}&({\bf s}_{k+1})|a_k=a,e_k=e,x_k=x\}
\nonumber\\
  =& \EE\{V_{\pi^*}(\clip_B(e - {c_k}),x_{k+1})|{|a_k=1}, x_k=x \} \nonumber\\
  =& a \EE\{V_{\pi^*}(\clip_B(e-{c_k)},x_{k+1}) {|a_k=1}\}   \nonumber\\
   &+ (1-a) \EE\{V_{\pi^*}(\clip_B(e-{c_k}),x_{k+1}) {|a_k=0}\}.
\label{EqCondV}
\end{align}
Joining \eqref{EqBellman}, \eqref{EqCondrk} and \eqref{EqCondV}, {and using as3)},
\begin{align}
V_{\pi^*}({\bf s}) &= \max_{a}\{a x W(e,x)  \\
        &+     a \gamma \EE\{V_{\pi^*}(\clip_B(e-{c_k}),x_{k+1})
                             {|a_k=1} \}  \nonumber\\
        &+ (1-a) \gamma \EE\{V_{\pi^*}(\clip_B(e-{c_k}),x_{k+1})
                             {|a_k=0} \} \}.    \nonumber
\label{EqBellman2}
\end{align}
Defining the threshold function
\begin{align}
\mu(e) = \gamma(\EE\{V_{\pi^*}(\clip_B(e-{c_k}),x_{k+1})
                               {|a_k=0}\}  \nonumber\\
              - \EE\{V_{\pi^*}(\clip_B(e-{c_k}),x_{k+1})
                               {|a_k=1}\}).
\end{align}
the optimal policy is $a^* = \pi^*(e,x) = u( x W(e,x) - \mu(e))${, which} is equivalent to \eqref{Eq.opt.a_k.Stat}

{Note also} that \eqref{EqBellman2} can be written as
\begin{align}
V_{\pi^*}({\bf s}) =& \gamma \EE\{V_{\pi^*}(\clip_B(e-{c_k}),x_{k+1})
                                  {| a_k=0}  \}
	\nonumber\\
	&+ (x W(e,x) - \mu(e))^+.
\label{EqBellman3}
\end{align}

Defining $\lambda(e)$ as in \eqref{EqDefLambda}, we get \eqref{Eq.Mu_Stat} and \eqref{Eq.Lambda_Stat}.

\section*{Appendix B: Transition probability matrix }
\label{TransitionProb}

The entries of the transition matrix ${\bf P}$ for an arbitrary transmission policy based on a generic threshold function $\mu(e)$ [cf. \eqref{Eq.trans_prob}] can be found (using the abbreviated notation $P\{j|i,\ldots\}$ instead of $P\{e_k =j|e_{k-1}=i,\ldots\}$) as
\begin{align}
p_{ij} =& P\left\{j|i,
            x_{k-1} \ge \frac{\mu(i)}{W(i)}\right\}
        \cdot \left(1-F_X\left(\frac{\mu(i)}{W(i)}\right)\right) \nonumber\\
        &+ P\left\{j|i,x_{k-1} < \frac{\mu(i)}{W(i)}\right\}
         \cdot F_X\left(\frac{\mu(i)}{W(i)}\right)                \nonumber\\
       =&\small{
         \left[\begin{array}{ll}
             (1 - F_{c_1}(i))(1-F_X) + (1-F_{c_0}(i)) F_X      & j=0   \\
             P_{c_1}(i-j)(1-F_X) + P_{c_0}(i-j) F_X    & 0<j<B \\
             F_{c_1}(i-B)(1-F_X) + F_{c_0}(i-B) F_X  & j=B
               \end{array}
         \right.
         }
         \label{Eqpij}
\end{align}
where $P_{c_0}$ and $P_{c_1}$ are the {conditional} probability mass functions of $c$ {given actions $a=0$ and $a=1$}, respectively, and $F_{c_0}$ and $F_{c_1}$ the respective cumulative {conditional} probability functions. With some abuse of notation, we have abbreviated $F_X=F_X(\mu(i)/W(i))$.


\section*{Appendix C: Derivation of the stochastic algorithm}
\label{RobbinsMonro}

Let us first define functions
\begin{align}
\alpha(e) = \EE\{\lambda(\clip_B(e-{c_{0,k}}))\} \\
\beta(e)  = \EE\{\lambda(\clip_B(e-{c_{0,k}-\ctk}))\}.
\end{align}
{The solution} \eqref{Eq.opt.a_k.Stat} of the MDP {can then} be written as
\begin{align}
a_k = u(W(e_k) x_k-\gamma(\alpha(e_k)-\beta(e_k))),
\label{Eq.opt.ak2}
\end{align}
\begin{align}
\lambda(e)
   &= \gamma \alpha(e) + \EE\{(W(e,{\bf z}_k) x_k- \gamma(\alpha(e)-\beta(e)))^+ \}.
\label{Eq.Lambda_Stat}
\end{align}

To derive the proposed algorithm as an instance of the Robbins-Monro algorithm, we represent functions in vector notation. Accordingly, we define $\boldsymbol{\lambda} = (\lambda(0),\lambda(1),\ldots,\lambda(B))^\intercal$, and let $\boldsymbol{\omega}$, $\boldsymbol{\alpha}$, $\boldsymbol{\beta}$ be the corresponding vectorizations of $W(e)$, $\alpha(e)$ and $\beta(e)$. We {also} define the vector of success indices
\begin{align}
{\bf w}_{c} = (u(0-c), u(1-c), \ldots, u(B-c))^\intercal
\end{align}
{and} the transformation $\boldsymbol{\lambda}' = {\bf T}_c \boldsymbol{\lambda}$, such that $\lambda'_i = \lambda_{\clip_B(i-c)+1}$. Then, we can write
\begin{align}
\label{DefOmega}
\boldsymbol{\omega}  &= \EE\{{{\bf w}_{c_{0,k}+\ctk}}\}                        \\
\boldsymbol{\alpha}  &= \EE\{{\bf T}_{c_{0,k}} \boldsymbol{\lambda}\}   \\
\boldsymbol{\beta}   &= \EE\{{\bf T}_{c_{0,k}+\ctk} \boldsymbol{\lambda}\}   \\
\boldsymbol{\lambda} &= \gamma \boldsymbol{\alpha}
   + \EE\{(\boldsymbol{\omega} x - \gamma(\boldsymbol{\alpha}-\boldsymbol{\beta}))^+ \}.
\label{DefLambdaVector}
\end{align}
Now, defining vector
\begin{align}
{\bf v} = \left(\boldsymbol{\omega}^\intercal,
                \boldsymbol{\alpha}^\intercal,
                \boldsymbol{\beta}^\intercal,
                \boldsymbol{\lambda}^\intercal \right)^\intercal
\end{align}
and matrices
\begin{align}
{\bf M}_{c_0,c_1} = \left(
\begin{array}{llll}
	 -{\bf I}      &{\bf 0}       & {\bf 0}  & {\bf 0}        \\
	 {\bf 0}       &-{\bf I}      & {\bf 0}  & {\bf T}_{c_0}  \\
	 {\bf 0}       &{\bf 0}       & -{\bf I} & {\bf T}_{c_1}  \\
	 {\bf 0}       &\gamma{\bf I} & {\bf 0}  & -{\bf I}
\end{array}
\right)
\end{align}
\begin{align}
{\bf N}_x = \left(
\begin{array}{llll}
	 {\bf 0}   & {\bf 0}         & {\bf 0}       & {\bf 0}  \\
	 {\bf 0}   & {\bf 0}         & {\bf 0}       & {\bf 0}  \\
	 {\bf 0}   & {\bf 0}         & {\bf 0}       & {\bf 0}  \\
	 x {\bf I} & -\gamma {\bf I} & \gamma{\bf I} & {\bf 0}
\end{array},
\right)
\end{align}
eqs. \eqref{DefOmega} to \eqref{DefLambdaVector} are equivalent to
\begin{equation}
\EE\left\{\left({\bf N}_x {\bf v} \right)^+
              + {\bf M}_{{c_{0,k},c_{0,k}+\ctk}}{\bf v}
              + \overline{\bf w}_{{c_{0,k}+\ctk}}
    \right\} = 0
    \label{eq:ob_robbins}
\end{equation}
where ${\overline{\bf w}_{c} = ({\bf w}_c^\intercal}, {\bf 0}^\intercal, {\bf 0}^\intercal, {\bf 0}^\intercal)^\intercal$. The Robbins-Monro algorithm that solves \eqref{eq:ob_robbins} then becomes \cite{yin2003stochastic}
\begin{equation}
{\bf v}_{k+1} = {\bf v}_k
              + \eta_k
                \left(\left({\bf N}_k {\bf v}_k \right)^+
                    + {\bf M}_k {\bf v}_k + \overline{\bf w}_k \right),
\end{equation}
which is equivalent {to \eqref{AlgOmega}-\eqref{AlgLambda}.}

\bibliographystyle{IEEEtran}
\bibliography{osf}

\begin{IEEEbiography}[{\includegraphics[width=1in,height=1.80in,clip,keepaspectratio]{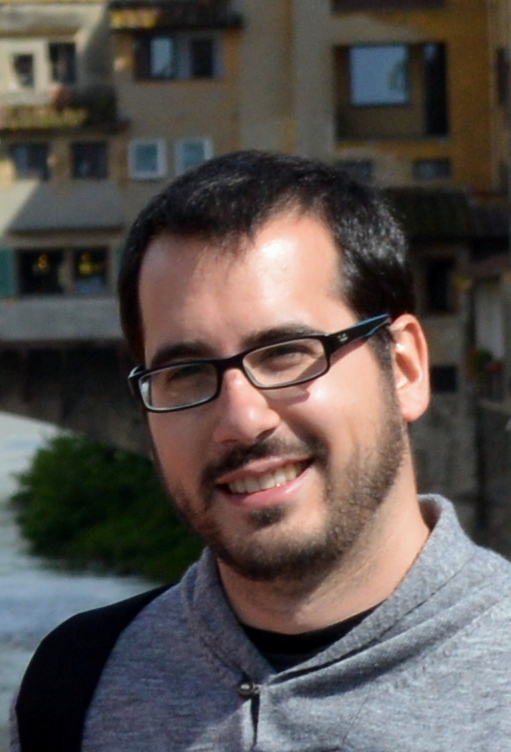}}]{Jesus Fernandez-Bes}
received the degree in Telecommunication Engineering (with honors) and Master in Multimedia y Communications from Universidad Carlos III de Madrid, Spain, in 2010 and 2012 respectively. 

He is currently working towards a Ph.D. degree at the Universidad Carlos III de Madrid, Madrid, Spain. His research is mainly focused on distributed estimation and intelligent energy management in wireless sensor networks.

\end{IEEEbiography}

\begin{IEEEbiography}[{\includegraphics[width=1in,height=1.80in,clip,keepaspectratio]{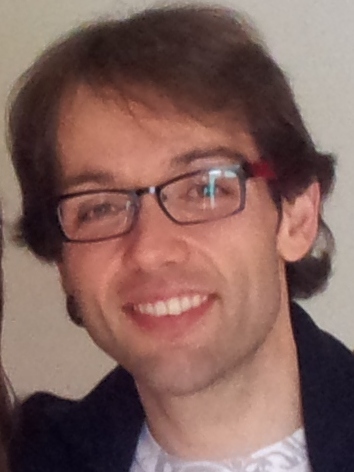}}]{Antonio G. Marques (SM'13)}
received the Telecommunications Engineering 
degree and the Doctorate degree (together equivalent to the B.Sc., 
M.Sc., and Ph.D. degrees in electrical engineering), both with highest 
honors, from the Carlos III University of Madrid, Spain, in 2002 and
 2007, respectively. In 2003, he joined the Department of Signal Theory
 and Communications, King Juan Carlos University, Madrid, Spain, where
 he currently develops his research and teaching activities as an 
Associate Professor. Since 2005, he has held different visiting 
positions at the University of 
Minnesota, Minneapolis. 

His research interests lie in the areas of communication theory,
 signal processing, and networking. His current research focuses on 
stochastic resource allocation for green and cognitive wireless networks, smart grids, 
nonlinear network optimization, and signal processing for graphs. Dr. 
Marques has served the IEEE in a number of posts (currently, he is an Associate Editor of the Signal Process. Letters) and his work has been awarded in several conferences and workshops.
\end{IEEEbiography}

\begin{IEEEbiography}[{\includegraphics[width=1in,height=1.80in,clip,keepaspectratio]{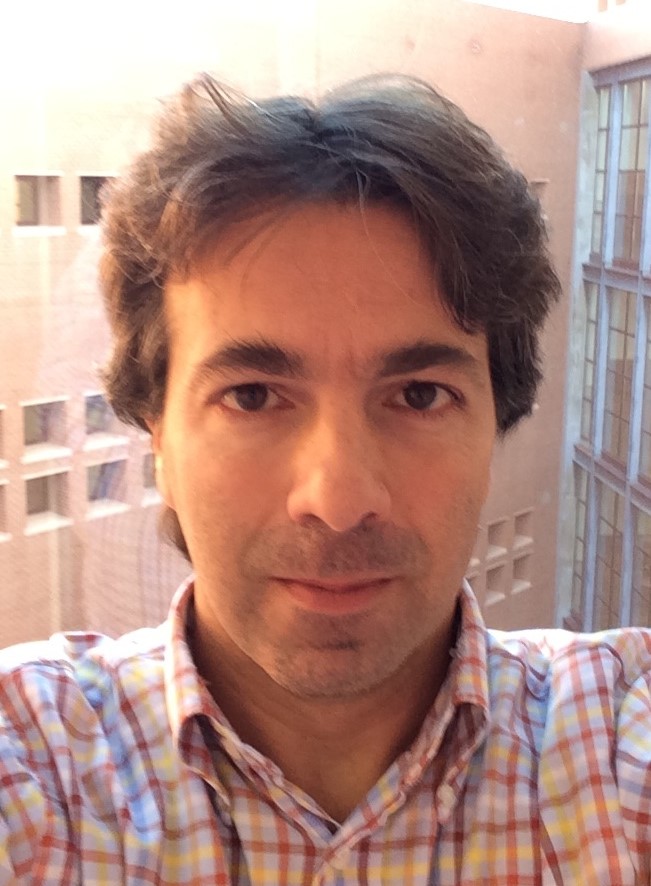}}]{Jes\'us Cid-Sueiro (M'95-SM'08)}
received the Degree in telecommunication engineering from Universidad de Vigo, Vigo, Spain, and the Ph.D. degree from Universidad Polit\'ecnica de Madrid, Madrid, Spain, in 1990 and 1994, respectively.

He is currently a Professor with the Department of Signal Theory and Communications, Universidad Carlos III de Madrid, Legan\'es. His current research interests include machine learning, Bayesian methods, computational intelligence and their applications in sensor networks, big data and signal processing.
\end{IEEEbiography}
\vfill

\end{document}